\documentclass[useAMS,usenatbib]{mn2e}

\usepackage{graphicx}

\newcommand{\Msun}{$M_{\odot}$}
\newcommand{\mas}{mas}
\newcommand{\may}{mas\,yr$^{-1}$}

\newcommand{\kms}{km\,s$^{-1}$}
\newcommand{\vs}{$v \sin i$}
\newcommand{\teff}{$T_{\rm eff}$}
\newcommand{\lgg}{$\log\,{g}$}

\title[Magnetism and chemical peculiarities in HD 72106]{Magnetic fields and chemical peculiarities of the very young intermediate-mass binary system HD 72106}

\author[Folsom et al.]{C.P. Folsom$^{1,2,3}$\thanks{E-mail: cpf@arm.ac.uk}, G.A. Wade$^{1}$, O. Kochukhov$^{4}$, E. Alecian$^{1}$, C. Catala$^{5}$, 
\newauthor
S. Bagnulo$^{3}$, T. B\"ohm$^{6}$, J.-C. Bouret$^{7}$, J.-F. Donati$^{6}$, J. Grunhut$^{1,2}$, 
\newauthor
D.A. Hanes$^{2}$, and J.D. Landstreet$^{8}$\\
$^{1}$Department of Physics, Royal Military College of Canada, P.O. Box 17000, Station `Forces', Kingston, Ontario, Canada, K7K 7B4\\
$^{2}$Department of Physics, Engineering Physics \& Astronomy, Queen's University, Kingston, Ontario, Canada, K7L 3N6 \\
$^{3}$Armagh Observatory, College Hill, Armagh Northern Ireland BT61 9DG\\
$^{4}$Department of Astronomy and Space Physics, Uppsala University, 751 20 Uppsala, Sweden \\
$^{5}$Observatoire de Paris, LESIA, 5 place Jules Janssen, F-92195 Meudon Cedex, France \\ 
$^{6}$Laboratoire d'Astrophysique, Observatoire Midi-Pyr\'en\'ees, 14 avenue Edouard Belin, F-31400 Toulouse, France\\
$^{7}$Laboratoire d'Astrophysique de Marseille, Traverse du Siphon, BP8-13376 Marseille Cedex 12, France\\
$^{8}$Physics \& Astronomy Department, The University of Western Ontario, London, Ontario, Canada, N6A 3K7 }

\begin{document}

\date{Received: someday; Accepted: one can dream}

\pagerange{\pageref{firstpage}--\pageref{lastpage}} \pubyear{2008}

\maketitle

\label{firstpage}

\begin{abstract}
The recently discovered magnetic Herbig Ae and Be stars may provide qualitatively new 
information about the formation and evolution of magnetic Ap and Bp stars. 
We have performed a detailed investigation of one particularly interesting 
binary system with a Herbig Ae secondary and a late B-type primary possessing a strong, globally
ordered magnetic field.  Twenty high-resolution Stokes $V$ spectra of the system were obtained 
with the ESPaDOnS instrument mounted on the CFHT. In these observations we see clear 
evidence for a magnetic field in the primary, but no evidence for a magnetic field in the secondary. 
A detailed abundance analysis was performed for both stars, 
revealing strong chemical peculiarities in the primary and normal chemical abundances in the secondary.  
The primary is strongly overabundant in Si, Cr, and other iron-peak elements, as well as Nd, and underabundant in He. 
The primary therefore appears to be a very young Bp star. In this context, line profile variations of the primary suggest
non-uniform lateral distributions of surface abundances. 
Interpreting  the $0.63995 \pm 0.00009$ day variation period of the Stokes $I$ and $V$ profiles as the
rotational period of the star, we have modeled the magnetic field  
geometry and the surface abundance distributions of Si, Ti, Cr and Fe using Magnetic Doppler
Imaging. We derive a dipolar geometry of the surface magnetic field,  
with a polar strength $B_{\rm d}=1230$~G and an obliquity $\beta=57\degr$. The distributions Ti, Cr and Fe are  
all qualitatively similar, with an elongated patch of enhanced abundance situated near the positive magnetic pole.  
The Si distribution is somewhat different, and its relationship to the magnetic field geometry less clear. 

\end{abstract}


\section{Introduction}
Strong, globally organised magnetic fields have recently been reported in a few 
Herbig Ae and Be (HAeBe) stars \citep{Donati1997-major, Hubrig2004-HAeBe, 
Wade2005-HAeBe_Discovery,Wade2007-HAeBe_survey,Catala2007-HD190073,Alecian2008-HD200775,Alecian2008-ClusterHAeBeLetter}.  HAeBe stars are 
pre-main sequence stars of intermediate-mass which evolve to become main sequence A and B stars.  
HAeBe stars have A or B spectral classes, display emission lines and infrared excesses, and 
are usually found with nebulosity nearby \citep{Vieira2003-HAeBe_ID}.  
The detection of magnetic fields in HAeBe stars is of particular interest 
as it hints at a connection to the main sequence magnetic, chemically 
peculiar Ap and Bp stars.  

Ap and Bp stars display strong magnetic fields, 
with typical strengths of a few hundred gauss up to a few tens of kilogauss, 
and globally ordered geometries of approximately dipolar topology.  These stars also display strong and distinctive chemical 
peculiarities, particularly overabundances of Si and iron peak elements, occasionally in excess of 2 dex, and 
even greater overabundances of some rare earth elements.  

The source of the magnetic fields observed in Ap/Bp stars is not well understood.  
The two major competing field origin theories propose, on the one hand, that the field is a relic 
of an earlier stage of stellar evolution, now frozen into the plasma of the star 
or, on the other hand, that the field is generated contemporaneously by a dynamo.  
Also unknown are the details of the formation of the 
observed chemical peculiarities.  Chemical peculiarities are believed to be the 
result of atomic diffusion leading to a chemically stratified stellar 
envelope \citep{Michaud1970-diffusion,Michaud1981-diffusion_magneticApBp}.  
However, the details of this process, particularly for individual elements,  
and the impact of magnetic fields are not fully understood \citep[e.g.][]{AlecianG2007-el-distributions-atmos}.  
In this context it is of great interest to identify and characterize the evolutionary progenitors of Ap/Bp stars, 
as these objects can provide critical information on both the origin of 
magnetic fields and the formation of chemical peculiarities.

The recently discovered magnetic HAeBe stars have been proposed to be 
pre-main sequence progenitors of Ap/Bp stars \citep{Wade2005-HAeBe_Discovery}.  Thus a detailed investigation of 
these stars can shed light on these questions, as well as potentially provide some 
information about magnetic braking and magnetospheric accretion in intermediate-mass stars.

\section{HD 72106}
\label{HD 72106 intro}

HD 72106 is a double star system in the constellation Vela, with a 0.805 arcsec separation between components \citep{Hipparcos1997}. 
The brighter star, HD 72106A (`the primary'), was identified as a magnetic star by \citet{Wade2005-HAeBe_Discovery}, 
while the fainter star, HD 72106B (`the secondary'), was identified as a HAeBe star by \citet{Vieira2003-HAeBe_ID}.
The HD 72106 system was observed by Hipparcos \citep{Hipparcos1997}, 
and included in the recent re-reduction of Hipparcos data 
by \citet{van_Leeuwen2007-Hipparcos_book,van_Leeuwen2007-Hipparcos_validation}
who found a parallax of $3.60 \pm 1.14$ \mas,
placing the system at a heliocentric distance of $278^{+129}_{-67}$ pc.
The system was identified as having an infrared excess, based on IRAS 
(the Infrared Astronomical Satellite) data, by \citet{Oudmaijer1992-IRAS+HD72106}.  
\citet{Torres1995-IRAS_TTauri+HD72106} observed HD 72106 and
noted the presence of emission in the H$\alpha$ Balmer line of the combined spectrum of the system.  
\citet{Schutz2005-IR_spec} obtained infrared spectroscopy of HD 72106 system and examined 
circumstellar abundances of a number of dust grain species.  

\citet{Vieira2003-HAeBe_ID} associated the system with the Gum Nebula star forming region. 
They observed weak H$\alpha$ emission and a small contribution from dust in the spectral energy distribution of HD 72106B.  
This, they hypothesized, was due to HD 72106B being an evolved HAeBe star that has cleared most of its 
circumstellar envelope. 

The discovery of a magnetic field in in HD 72106A by \citet{Wade2005-HAeBe_Discovery} was
based on spectropolarimetry from FORS1 (FOcal Reducer/low dispersion Spectrograph) 
at the Very Large Telescope and from ESPaDOnS (Echelle SpectroPolarimetric Device for the 
Observation of Stars) at the Canada France Hawaii Telescope.  
A longitudinal field of $195 \pm 40$ G (i.e. 4.9$\sigma$) was deduced from the FORS1 spectrum. 
No significant magnetic field was detected in the secondary (65 $\pm$ 55 G was observed).
Although the average longitudinal field in the ESPaDOnS spectrum of the primary was consistent with zero, 
a clear circular polarization signature was detected with a very high degree of significance, 
unambiguously indicating the presence of a surface magnetic field.  

A more conservative re-analysis of the same FORS1 data by \citet{Wade2007-HAeBe_survey} 
did not confirm the longitudinal magnetic field detection at $3\sigma$ confidence.  
\citet{Wade2007-HAeBe_survey} reported longitudinal fields in HD 72106A of $166 \pm 70$ G from Balmer lines 
and $-11\pm 91$ G from metallic lines, 
and fields in HD 72106B of $52 \pm 90$ G from Balmer lines and $3 \pm 122$ G from metallic lines.  
However they stressed that ESPaDOnS observations show a clear circular polarization signal, 
implying that HD 72106A is definitely magnetic.  

Due to its young age, probable binarity, and magnetic properties, 
the HD 72106 system is a compelling target for further study.  
While other binary systems containing a magnetic intermediate-mass star are known, 
the apparently very young age and wide visual separation of HD 72106 make this system uniquely interesting. 
We therefore obtained 20 high-resolution spectropolarimetric (Stokes $I$ and $V$) observations of the system, 
with the aim of studying the system's binarity, rotational properties, magnetic field, chemical abundances, 
and surface abundance distributions.

\section{Observations}
Observations used in this paper were obtained with the ESPaDOnS 
instrument on the 3.6\,m Canada-France-Hawaii Telescope (CFHT).  
ESPaDOnS\footnote{ See  { \scriptsize http://www.cfht.hawaii.edu/Instruments/Spectroscopy/Espadons/} for details.}
 is a high resolution echelle spectropolarimeter, 
with a resolving power of about 65000 in spectropolarimetric mode, and nearly continuous wavelength 
coverage from 370 to 1050 nm.  This instrument uses a fiber fed design, with a polarimeter 
module mounted at the Cassegrain focus of the CFHT and an echelle spectrograph located 
in the Coud\'e room.  

Twenty Stokes $I$ and $V$ spectra were obtained over a period of two years, 
and are summarized in Table \ref{table_obs}.
``Normal'' readout in spectropolarimetric mode was used, and the atmospheric dispersion 
corrector was employed for all observations. 
Magnetic and non-magnetic standard stars were observed as part of this campaign, and produced 
results in excellent agreement with the literature. 

\begin{table*}
\centering

\caption{Log of observations of the HD 72106 system obtained with ESPaDOnS. 
Integration times refer to the complete set of 4 sub-exposures 
used to produce each Stokes $V$ spectrum \citep{Donati1997-major}. Peak signal-to-noise 
ratios are reported for the original reduced spectra with 1.8 \kms\, spectral pixels. 
Magnetic measurements for combined observations refer to results for the reconstructed spectra of the primary.
The letter D in the LSD detection 
column indicates that the Stokes $V$ LSD profile is inconsistent with the null field hypothesis at the 
99.999\% confidence limit, indicating the presence of a magnetic field. 
The letter N indicates that the inconsistency was at the 99.9\% limit or less, 
and hence cannot be take as strong evidence for the detection of a magnetic field. 
The letters MD indicate a `marginal' detection, with a value between 99.9\% and 99.999\%, 
and are considered suggestive of a magnetic field, but not conclusive. \citep{Donati1992-ZeemanDI,Donati1997-major}.
Note that the observation from Feb. 2005 was first published by \citet{Wade2005-HAeBe_Discovery}.  }

\begin{tabular}{cccccccrc}
\hline \hline \noalign{\smallskip}
UT Date    &HJD         & Component & Integration & Peak S/N & $B_z$           & LSD        \\           
           &(-2 450 000)& (HD 72106)&  Time (s)   &    $I$   & (G)             & Detection  \\
 \noalign{\smallskip} \hline \noalign{\smallskip}
22 Feb. 05 & 3423.9248  & A\&B   & 2400         &  201     &   228  $\pm$   50 & D \\
09 Jan. 06 & 3745.02967 & A\&B   & 2400         &  219     &   345  $\pm$   42 & D \\
11 Jan. 06 & 3747.02034 & A\&B   & 3200         &  143     &   261  $\pm$   76 & MD\\
12 Jan. 06 & 3747.99629 & A      & 1200         &  149     &   -13  $\pm$   44 & MD\\
12 Jan. 06 & 3748.01496 & B      & 1200         &   76     &     2  $\pm$  168 & N \\
11 Feb. 06 & 3777.87860 & A\&B   & 2000         &  238     &   231  $\pm$   40 & D \\
11 Feb. 06 & 3777.95149 & A\&B   & 2400         &  253     &   124  $\pm$   37 & D \\
12 Feb. 06 & 3778.86172 & A\&B   & 2400         &  282     &   282  $\pm$   34 & D \\
13 Feb. 06 & 3779.87202 & A\&B   & 2400         &  184     &   170  $\pm$   53 & D \\
13 Feb. 06 & 3779.98127 & A\&B   & 2400         &  128     &   157  $\pm$  156 & N \\
02 Mar. 07 & 4161.77282 & A\&B   & 2400         &  254     &   350  $\pm$   36 & D \\
02 Mar. 07 & 4161.80256 & A\&B   & 2400         &  265     &   297  $\pm$   34 & D \\
02 Mar. 07 & 4161.90282 & A\&B   & 2400         &  297     &   214  $\pm$   30 & D \\
03 Mar. 07 & 4162.85713 & A\&B   & 2400         &  208     &   246  $\pm$   44 & D \\
04 Mar. 07 & 4163.83561 & A\&B   & 2400         &  322     &   202  $\pm$   27 & D \\
05 Mar. 07 & 4164.84650 & B      & 3200         &  209     &   -51  $\pm$   55 & N \\
05 Mar. 07 & 4164.88387 & A      & 2400         &  248     &   320  $\pm$   23 & D \\
05 Mar. 07 & 4164.90961 & A\&B   & 1600         &  277     &   374  $\pm$   32 & D \\
09 Mar. 07 & 4168.85791 & A\&B   & 2400         &  283     &   304  $\pm$   33 & D \\
09 Mar. 07 & 4168.90947 & A\&B   & 2400         &  272     &   252  $\pm$   34 & D \\
\noalign{\smallskip} \hline \hline

\end{tabular}

\label{table_obs}

\end{table*}

With careful guiding and monitoring, observations of the individual 
components of HD 72106 were acquired on a few nights with particularly good seeing.  
This was challenging, however, as the double star system has a separation of only 0.805 arcsec 
and ESPaDOnS has a pinhole with a diameter of 1.6 arcsec.
On most nights the atmospheric conditions were not good enough for us to 
resolve the individual components of the double star system.  In these cases the 
photocenter of the system was observed, providing combined spectra of the system.  

The observations were reduced with Libre-ESpRIT \citep[][and in preparation]{Donati1997-major}.  
This is a nearly-automatic dedicated data reduction 
package for ESPaDOnS, which performs 
a complete calibration and optimal spectrum extraction.  
Continuum normalization of the resulting unnormalized spectra was performed manually, 
first by computing a running average on each order of an observed spectrum, 
then selecting points in the continuum (i.e. maxima) of the smoothed order, 
and fitting a polynomial (typically of degree 4 or 5) through the selected points.  
The original observation was then 
divided by the polynomial to produce a normalized spectral order, 
and the process was repeated for each order in the echelle spectrum.  
Similarly, Stokes $V$ spectra were normalized by the Stokes $I$ continuum polynomial, producing a $V/I_c$ spectrum.

\section{Spectrum Reconstruction}
\label{Spectrum Reconstruction}
\subsection{Procedure}
\begin{figure*}
\centering
\includegraphics[width=3.0in]{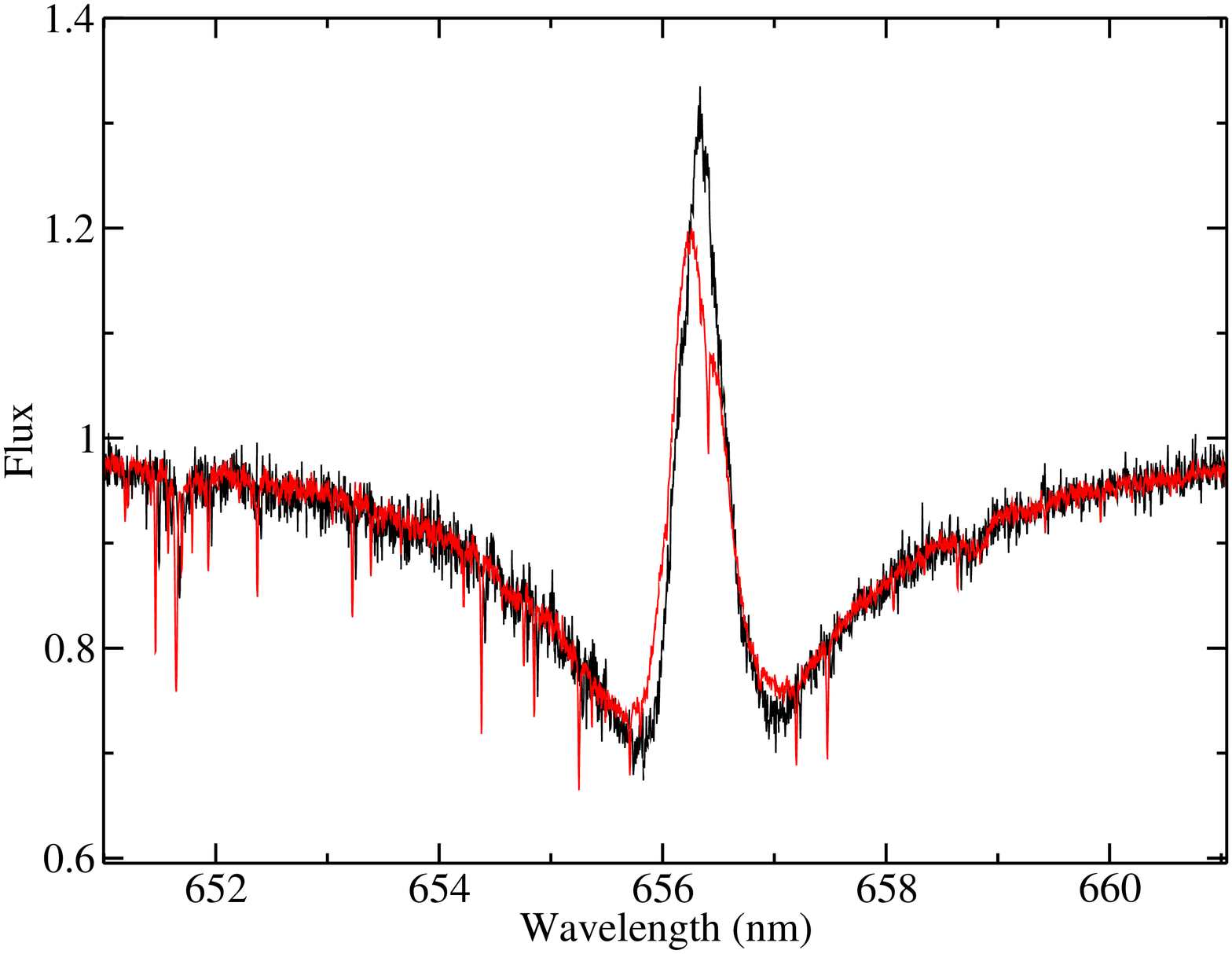}
\includegraphics[width=3.0in]{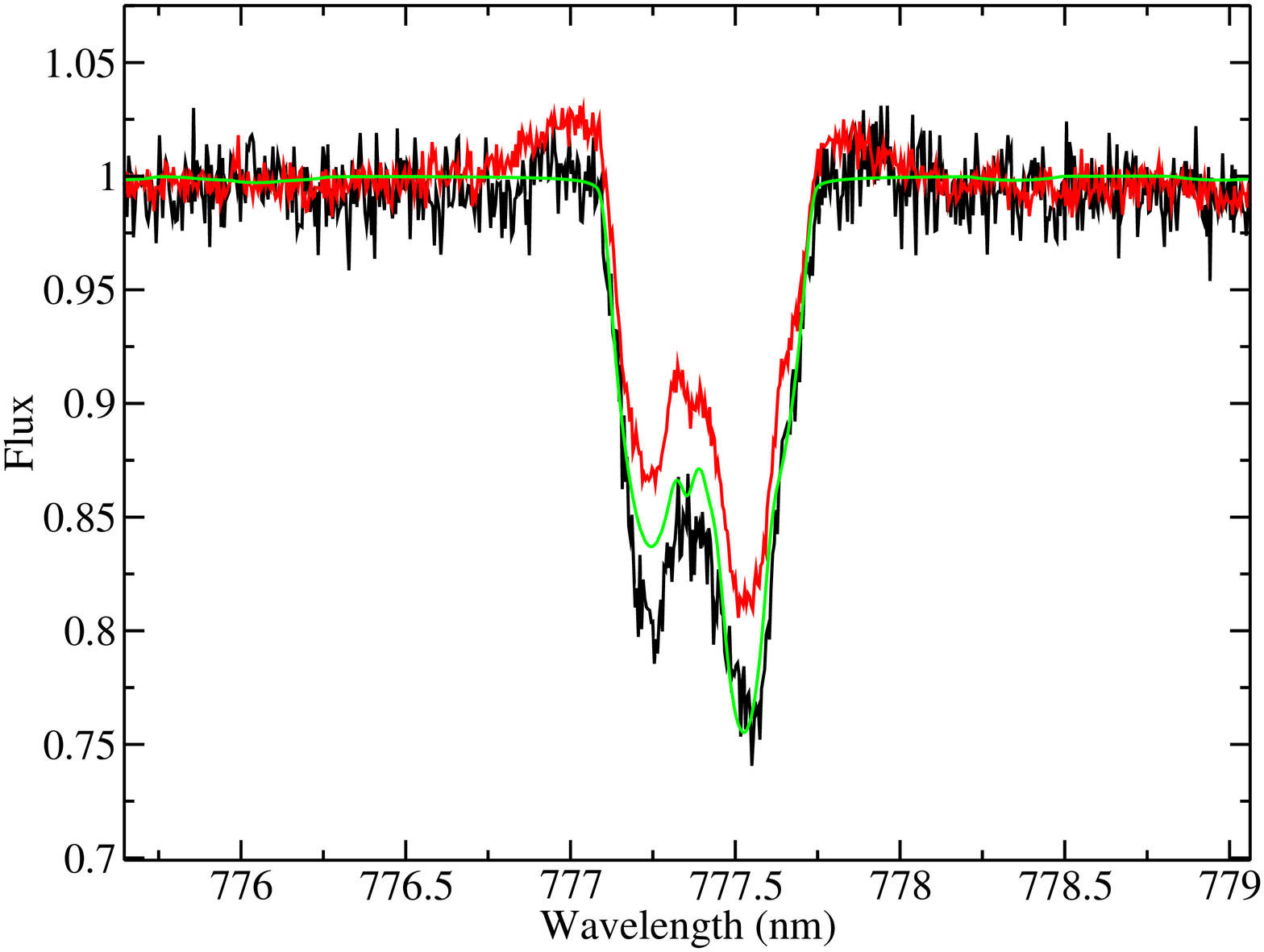}
\caption{Emission and variability in the H$\alpha$ Balmer line and the O {\sc i} 7771 \AA\, 
triplet of HD 72106B (the HAeBe secondary), shown for the two single spectra of HD 72106B.  
The smooth line in the O {\sc i} 7771 \AA\, panel is an LTE synthetic profile calculated assuming solar abundances. 
Strong, variable emission can be seen in H$\alpha$, while mostly we see only emission infilling in O {\sc i} 7771 \AA.
No other lines in the spectrum of the secondary show any evidence of variability. }
\label{emission-fig}
\end{figure*}

The two individual spectra of HD 72106A show strong metallic lines with significant variability. 
In the two observations of HD 72106B on its own we see no metallic absorption line variability, 
down to the level of the noise in our observations.  
However, we do see emission in two lines of HD 72106B: 
clear, variable, emission is present in H$\alpha$, and a small amount of 
emission in the O {\sc i} 7771 \AA\, triplet, as shown in Figure \ref{emission-fig}.
Thus we confirm the spectroscopic properties of the secondary leading to its classification as a HAeBe star.
The emission in HD 72106B varies detectably from night to night.  
Stronger variations are observed on timescales of months to years.  
No detectable change is observed between observations obtained on the same night. 
A period analysis of the emission line variations yields no significant periodicity in H$\alpha$. 

As a consequence of the $\sim$0.8 arcsec angular separation between the components of HD 72106, 
the majority of our spectra of HD 72106A \& B are of the combined 
light from the system rather than light from the individual components.  The analysis of observed spectra 
is much more tractable when one is dealing with light from only one object.  
Thus we attempted to reconstruct the spectrum of one star from the combined spectra of the system.  

As the spectrum of the secondary does not appear to vary (outside of emission lines), 
we can subtract it from the combined spectra and reconstruct the spectrum of the variable primary.  
We begin by modeling the normalized flux $I_{\lambda,T}$ at any point in 
the combined spectrum according to:

\begin{equation}
\label{adding spectra}
I_{\lambda,T} L_T = I_{\lambda,1} L_1 + I_{\lambda,2} L_2,
\end{equation}
where $I_{\lambda,1}$ and $I_{\lambda,2}$ refer to the normalized 
flux spectrum of the primary and secondary respectively at wavelength $\lambda$, 
and $L_1$ and $L_2$ refer to the luminosity of the primary and secondary respectively.  
Thus the product $I_{\lambda,1} L_1$ gives the total flux of the primary.  
$I_{\lambda,T}$ and $L_T$ are the observed normalized flux spectrum of the system 
and the total luminosity of the system respectively.  The total luminosity can be written as 
$L_T = L_1 + L_2$.  It is elementary to rearrange Eq.~(\ref{adding spectra}) 
to obtain the normalized spectrum of the primary, $I_{\lambda,1}$:
\begin{equation}
I_{\lambda,1} = I_{\lambda,T} \left(1 + \frac{L_2}{L_1}\right) - I_{\lambda,2}\frac{L_2}{L_1} .
\label{spec-subtract}
\end{equation}
Thus the normalized spectrum of the primary can be reconstructed from an observed spectrum of the system ($I_{\lambda,T}$) 
and a spectrum of the secondary on its own ($I_{\lambda,2}$).
Formally the luminosity ratio should also depend on wavelength, however since the two components 
differ in temperature by only 2000~K \citep[see Section \ref{Fundimental Physical Properties}, or][]{Wade2007-HAeBe_survey} the wavelength 
dependence is not strong, and thus has been neglected. 
Nevertheless, the magnitudes used to calculate the luminosity ratio must approximately correspond to the 
wavelength range of interest in the observed spectrum.  
The magnitude difference between components ($0.62 \pm 0.02$ in the Tycho $V$ band)
is well known from Hipparcos and Tycho data \citep{Hipparcos1997,Fabricius2000-Hipp-rered}, 
thus the ratio of luminosities at $\sim 5000$~\AA\, is well established.  

\subsection{Quality Control}
To investigate the  importance of the assumption of a wavelength-independent luminosity  
ratio, we used ATLAS9 flux distributions corresponding to the  
atmospheric parameters of the primary and secondary stars (Table \ref{physical prop tab}).  
The synthetic flux distributions show that the monochromatic luminosity ratio varies by 15\% over the ESPaDOnS spectral range. 
However, since the reconstructed spectrum is essentially a difference spectrum, variations in the luminosity ratio correspond to variations in line depth
Thus a typical line with a depth of 10\% of the continuum, at the red or 
blue end of the spectrum, would have an inaccuracy of $\sim$1.5\% of the continuum, which is near the noise level in the observations.  
For most of the lines used in this study this systematic error is even smaller.  
Comparisons were performed using subtracted spectra calculated assuming different luminosity ratios, 
as well as comparisons of our reconstructed spectra to observations of the individual components.  
These showed that systematic errors due to our assumption of a wavelength independent luminosity ratio generally are 
well below the noise level in our spectra (with the exception of the Balmer lines, 
which were not included in our analyses). 

Stokes $V$ spectra of the primary were reconstructed by attributing the observed polarization 
signature to the primary and scaling the spectrum based the luminosity ratio of the stars.
The assumption that the polarization signal is due solely to the primary is quite reasonable 
as we see no sign of circular polarization in our spectra of just the secondary.
For the purposes of the subtraction, the higher S/N observation of the secondary (HJD 2454164.84650) was used.

Contamination of the individual spectra of the secondary by stray light  
from the primary is potentially a serious source of error in our analysis. 
However, in our observations this contamination appears to be minimal.  
Observations of the stars obtained on two different nights, under different seeing conditions, 
produce identical line profiles (in stable regions of the spectrum) at the level of the noise.  
If there were a significant amount of contamination, one would expect the contamination to vary with the seeing conditions. 
Thus the fact that we do not see any such variation suggests that the spectra are not detectably contaminated.  
A theoretical estimate of the fractional contamination in our spectra, based on the seeing and geometry of the observations, 
provides further support to this claim.  The stars were observed with the center of the target star on the edge 
of the pinhole (0.8 arcsec from the pinhole center) and the other star placed outside the pinhole, 
$\sim$1.6 arcsec away from the pinhole center.  
Assuming a Gaussian point spread function for both stars, with a full width at half maximum equal to the seeing, 
one can integrate through the pinhole to estimate the flux from each star in our observation.  
Normalizing the Gaussians by the relative luminosity of the stars, and using the worst seeing 
conditions for our observations (0.9 arcsec for the primary and 0.7 arcsec for the secondary), 
we obtain a theoretical worst-case contamination of 6\% (of the total observed flux) 
for the observation of the primary and 7\% (of the total observed flux) for the observation of the secondary.  
For contamination by a typical strong line (10\% of the continuum), a deviation of roughly 0.6\% of 
the continuum would be seen in the observation, which at about the noise level of our spectra.
Thus we conclude that such contamination is not a serious problem. 

\begin{figure}
\centering
\includegraphics[width=3.3in]{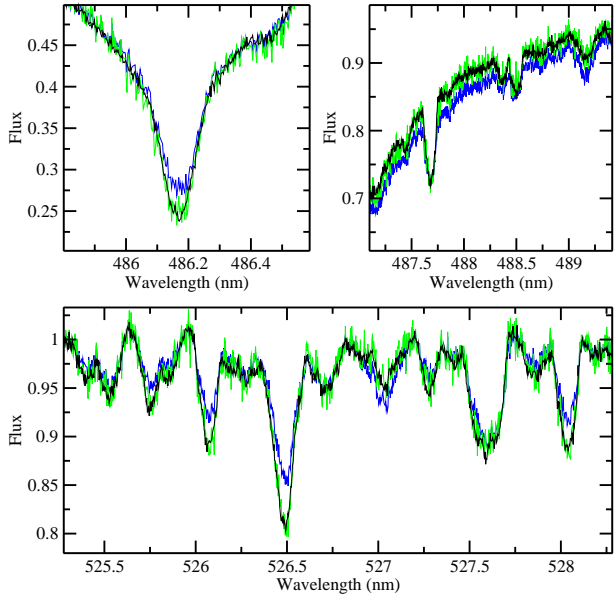}
\caption{Comparison of an observation of just HD 72106A (black, 5 March 2007), 
a combined observation of HD 72106A \& B (dark gray/blue, 11 January 2006), 
and the corresponding reconstructed spectrum of HD 72106A (light gray/green) 
obtained using the procedure described in Section \ref{Spectrum Reconstruction}.  
A good match of the reconstructed spectrum to the observation of the primary can 
be seen in the core (upper left) and wings (upper right) of H$\beta$, 
as well as in weakly variable metallic lines (lower panel).}
\label{spec-subtraction}
\end{figure}

Calculation of the spectrum of the primary from Eq.~(\ref{spec-subtract}) was performed for all our 
combined observations of the system, 
allowing us to reconstruct a sequence of 18 spectra of the primary, 
typical with a peak S/N ratio of 140 (at 5200~\AA).  
Comparisons of the reconstructed spectra to observations of the primary on its own show excellent concordance 
in the Balmer lines, and in weakly variable metallic lines, as shown in Figure \ref{spec-subtraction}.  
Furthermore, this procedure can be applied to recover the spectrum of the 
secondary (with a typical peak S/N of 90) if we use regions containing only weakly variable lines of the primary. 
Such experiments also yield good agreement with observations of the secondary on its own.  
Thus we conclude that the spectrum reconstruction is robust, 
and yields high-precision spectra of the primary component.

\section{Binarity and Evolutionary State}

\subsection{Binarity}
Determining whether the HD 72106 system is truly a binary or just an optical double star 
(i.e. an accidental conjunction of 2 stars at different distances along the line of sight)
is critical.  If the system is a coeval binary, this fact allows us to constrain the age of the 
primary much more accurately than would otherwise be possible.  Additionally, it suggests that both components 
formed from approximately the same material, making HD 72106 an interesting system 
from the point of view of stellar magnetic and chemical evolution.  

The system has a projected separation of 0.805 arcsec, 
and is thus fairly widely separated.  
Given the Hipparcos parallax of $3.60 \pm 1.14$ mas, this implies a 
minimum physical separation of $224 \pm 71$ AU.  

The Hipparcos observations were solved as a binary system producing a ``good quality'' solution for the system. 
This implies that the stars have the same parallax and 
proper motions, at the precision of the Hipparcos observations.  
Hipparcos finds a large proper motion in right ascension of  $-6.41 \pm 0.82$ \may\, 
and in declination of $7.97 \pm 1.23$ \may\, \citep{van_Leeuwen2007-Hipparcos_book}.
In the spectrum fitting procedure, described in Section \ref{Spectrum Synthesis},
we included heliocentric radial velocities for each star as a free parameter.  
From this fitting, we found identical radial velocities of $22 \pm 1$ \kms\, 
for both components.  Thus it appears that the stars are at the same point in space and moving together 
in three dimensions, 
strongly suggesting that they are in fact physically associated.  

\begin{figure}
\centering
\includegraphics[width=3.0in]{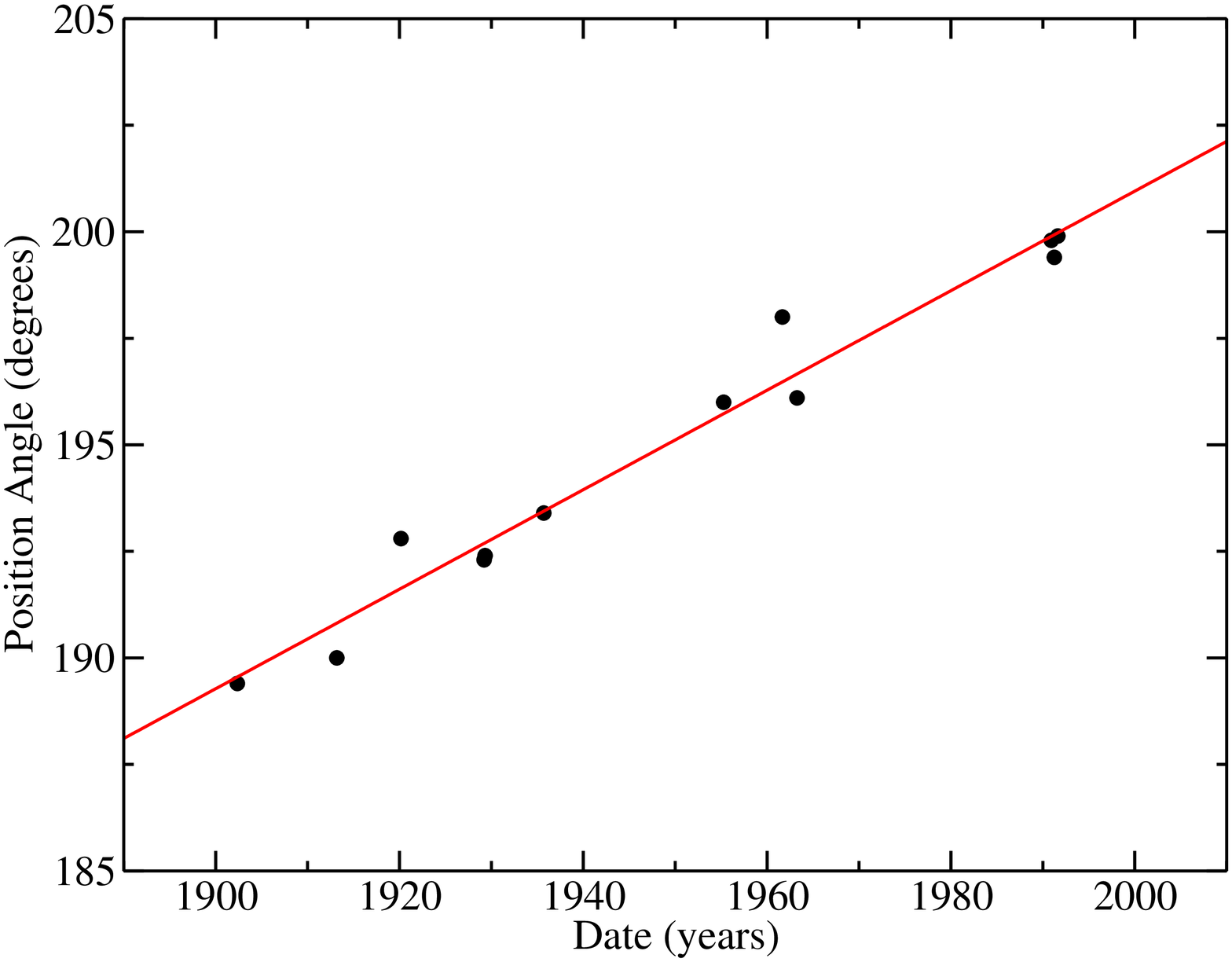}
\caption{Position angle measurements of the HD 72106, extracted from the WDS, 
together with their best fit line.  }
\label{position angles}
\end{figure}

Dr. Brian Mason at the United States Naval Observatory kindly provided a record of
separation and position angle observations from the Washington Double Star Catalog (WDS),
dating back to 1902 (private correspondence, Mason, 2008). These have been compiled 
from a number of sources in the literature, some of which are no longer readily accessible.  
These observations indicate no significant change in the separation of the components in 89.37 years, but show 
a clear and systematic increase in the position angle of the stars of $\sim$10.5$\degr$, illustrated in Figure \ref{position angles}.
This relative motion was not noted by \citet{Wade2007-HAeBe_survey}, presumably because 
they examined only a subset of these observations.  
The change in position angle occurs at a rate of $0.117 \pm 0.007\degr$~yr$^{-1}$. 
In contrast, there is no trend in the set of separation measurements, 
with an average separation of 0.786 arcsec and a standard deviation of 0.074 arcsec.
This kinematical behavior, together with the identical radial velocities 
and consistent parallaxes of the components, makes it high 
likely that the system is a true binary.  

For the system to be a true binary it must be gravitationally bound.  We have therefore verified 
that the observed properties are consistent with binarity.  
We assumed circular orbits, that the minimum possible separation is the true separation ($224 \pm 71$ AU), 
and used the evolutionary masses derived in Section \ref{Fundimental Physical Properties}.  
These assumptions are necessary, as we cannot determine the true 
separation and orbit of the binary without much more accurate observations 
over a much longer period of time.
In this geometry, the relative velocity of the stars is $4.1 \pm 0.7$ \kms\, 
and the orbital period is $1600 \pm 600$ years.  
This relative velocity is larger than observed, implying that the orbital plane is not parallel to the line of sight, 
or that the stars' physical separation is larger than we have assumed (which it almost certainly is).
If the orbital plane of the stars is in the plane of the sky, 
the orbital period implies that an observed change 
in position angle of $0.219 \pm 0.079\degr$~yr$^{-1}$ should be seen, 
which is roughly consistent with the WDS observed rate of change.  
Thus we conclude that a wide range of orbits are consistent with the observed radial velocities 
and position angle change.  One example of an orbit that is consistent with 
all observations is a circular orbit with a semimajor axis of 350 AU (implying a period of 3200 years), 
with an inclination of the orbital plane to the line of sight of $44.5\degr$, and a position angle of the orbital axis of $200\degr$.  
In this scenario, the two components would both be crossing the line-of-sight at the current epoch, yielding zero relative radial velocity. 
The change in position angle would be $0.112\degr$~yr$^{-1}$ and a apparent separation would be 0.807 arcsec.  
In this example, in 800 years the components would achieve their maximum difference in relative radial velocity of 2.3 km/s, 
and their maximum apparent separation of 1.151 arcsec. Of course, there is no reason to prefer this particular solution 
over the many others that are consistent with observations. 

When placed on the Hertzsprung-Russell (H-R) diagram, 
as discussed in Section \ref{Fundimental Physical Properties}, the stars are 
found to have positions that allow for a range of coeval solutions.  
Thus according to the H-R diagram positions, it is quite possible for 
the stars to have formed and evolved together, 
further supporting the binary hypothesis.  
 
Thus we conclude that HD 72106 is very likely a true binary system.  
The stars are at the same position in space, moving in the same 
(three dimensional) direction at the same velocity, allow for a wide range of gravitationally 
bound orbits, and appear to be consistent with coeval evolution.

\subsection{Fundamental Physical Properties}
\label{Fundimental Physical Properties}

Effective temperature and surface gravity for both components of HD 72106 were determined by
fitting Balmer lines.  Unfortunately, no intermediate-band photometry of the components of HD 72106 was available.  
The H$\alpha$ Balmer line in HD 72106B is clearly contaminated with emission, 
and hence was not considered in this analysis.  The other Balmer lines of both stars appear to be free 
of emission, and lines of each star can be well-fit with a single model atmosphere.  

\citet{Wade2005-HAeBe_Discovery} performed Balmer line fitting for both components of HD 72106 using 
observations from the FORS1. 
For the primary they found: \teff\, = $11000 \pm 1000$ K 
and $3.5\leq$ \lgg\, $\leq 4.5$, (best fit \teff\, = 11000 K \lgg\, = 4.0). 
For the secondary they found \teff\, = $8000 \pm 500$ and 
$4.0\leq$ \lgg\, $\leq 4.5$, (best fit \teff\, = 8000 K \lgg\, = 4.5).
We repeated the fitting procedure with the FORS1 spectra, 
using the method outlined above. 
Solar abundance ATLAS9 model atmospheres were used \citep{Kurucz1993-ATLAS9etc} to produce the synthetic Balmer 
lines and the model lines were convolved with a Gaussian instrumental profile 
of appropriate width to match the observations. 
We arrived at results identical 
to \citet{Wade2005-HAeBe_Discovery} for the primary. 
For the secondary 
we find the temperature range 7500 K to 9000 K and the \lgg\, range 4.0 to 4.5 more realistic, 
with a best fit value of 8000 K at \lgg = 4.5.  
These values from the FORS1 observations are also consistent with those determined from Balmer line 
fitting of our ESPaDOnS spectra.

In Section \ref{Spectrum Synthesis} we performed detailed spectrum synthesis 
of both components of HD 72106 and found that the atmospheric parameters 
of the secondary derived from Balmer lines are not compatible with the metallic line spectrum.
In particular, we were unable to simultaneously fit lines of Fe~{\sc i} and Fe~{\sc ii}, Cr~{\sc i} and Cr~{\sc ii}, 
and Ti~{\sc i} and Ti~{\sc ii}, suggesting a problem with the local thermodynamic equilibrium (LTE) ionization balance of these species.  
Including \teff\, and \lgg\, as free parameters in the spectrum fit allowed us to 
satisfactorily match the metallic line spectrum, provided that \teff\, $\sim$8750 K and \lgg\, $\sim$4.0.
While these values produce a somewhat poorer fit to the Balmer lines, the fit to metallic 
lines is improved substantially.  Hence we conclude that the higher temperature 
and lower \lgg\, provide a better description of the atmosphere of HD 72106B.

Including the full range of \teff\, and \lgg\, which provide acceptable fits to both 
the metallic line spectra and the Balmer lines of HD 72106B, 
we adopt  \teff\, = $8750 \pm 500$ K and \lgg\, = $4.0 \pm 0.5$. 

Using the Hipparcos parallax of the HD 72106 system 
and the Tycho magnitudes \citep{Fabricius2000-Hipp-rered} we determined the luminosities of both 
components of HD 72106.  To convert the observed Tycho $V$ magnitudes ($V_T = 9.00 \pm 0.01$ for the primary 
and $V_T = 9.62 \pm 0.02$ for the secondary) into a Johnson $V$ magnitudes ($V$)
we used the empirical relation \citep{Hipparcos1997}:
\begin{equation}
V = V_T + 0.09(B_T - V_T),
\end{equation}
where $B_T$ is the Tycho $B$ magnitude.  For the secondary, the bolometric 
correction for main sequence stars from \citet{Gray2005-Photospheres} was used, 
yielding a value of $0.01 \pm 0.06$ mag.  For the primary, 
the bolometric correction relation of \citet{Landstreet2007-surveyFORS1} 
was used, yielding a value of $-0.37 \pm 1.9$ mag.  This calibration is tailored specifically 
for magnetic chemically peculiar A and B type (Ap and Bp) stars.   
As will be shown, the primary has properties very similar to those of a Bp star, 
thus the calibration of \citet{Landstreet2007-surveyFORS1} is more appropriate. 
From this procedure we find that the luminosity of the primary  is $22\,^{+20}_{-10}\, L_\odot$, 
and the luminosity of the secondary is $9.2\,^{+8.2}_{-4.3} \, L_\odot$.
From the Stefan-Boltzmann equation we find identical 
radii, given our uncertainty, of $1.3 \pm 0.5 \, R_\odot$ for the primary and 
$1.3 \pm 0.5 \, R_\odot$ for the secondary (with $1\sigma$ error bars).  

\begin{figure}
\centering
\includegraphics[width=3.3in]{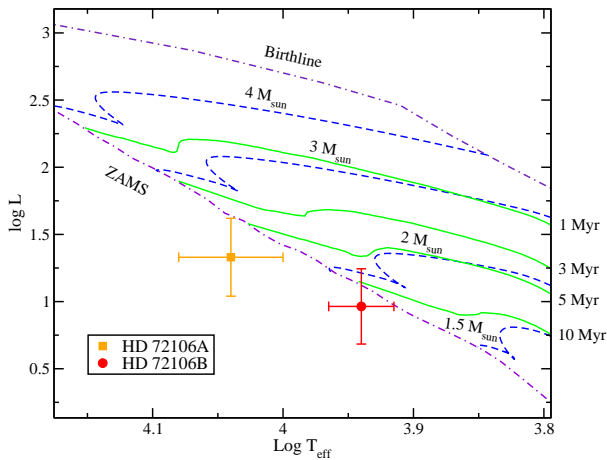}
\caption{An H-R diagram containing both components of the 
HD 72106 system.  Isochrones (solid lines), evolutionary tracks (dashed lines), 
and the zero age main sequence (ZAMS) line, are calculated from CESAM models \citep{Morel1997-CESAM}.
The birth line (for an accretion rate of $10^{-5}$ \Msun\, yr$^{-1}$)  is taken from \citet{Palla1993-PMS-Evol}. }
\label{h-r diagram}
\end{figure}

With the luminosity and effective temperature determined, we can place the stars on the 
H-R diagram, as shown in Figure \ref{h-r diagram}, and compare 
their positions with theoretical evolutionary tracks and isochrones.
Using the pre-main sequence evolutionary model calculations from CESAM \citep[version 2K;][]{Morel1997-CESAM} and the birthline of \citet{Palla1993-PMS-Evol}, 
we find the mass of the primary to be $2.4 \pm 0.3\, M_\odot$
and mass of the secondary to be $1.9 \pm 0.2\, M_\odot$ (with $1\sigma$ error bars).
From this we derive an evolutionary \lgg\, of  $4.6 \pm 0.3$ for the primary, 
and $4.5 \pm 0.3$ for the secondary, both of which are consistent with 
(although systematically higher than) the spectroscopic values.
We determine the binary's age to be between 6 and 13 Myr (measured from the birth line), 
based on the position of the secondary in the H-R diagram and assuming a 
protostellar accretion rate of $10^{-5}$ \Msun\, yr$^{-1}$.  
In doing this we have assumed, as most investigators do, that the presence of circumstellar material 
implies that the HAeBe secondary is on the pre-main sequence, and thus that
we can constrain its age with pre-main sequence evolutionary tracks.  
Given the derived admissible range of age and mass, the primary could 
still be on the pre-main sequence, reaching the zero-age main sequence (ZAMS) in less than 1 Myr.  
The most likely case (from `best fit' positions) is that 
the system is $\sim$10 Myr old, and the primary has just 
entered the main sequence, while the secondary is on its final approach to the ZAMS.  
In this case the primary would have spent $\sim$6 Myr on the main sequence, 
giving it a fractional age on the main sequence ($\tau$) of 0.01.  
In the oldest limiting case, the secondary is just reaching the ZAMS and the primary has 
been on the main sequence for $\sim$9 Myr giving it a fractional age of 0.015. 
Further observations, particularly a more accurate distance measurement, are necessary 
to more precisely determine the evolutionary status of the primary.  
Regardless, while HD 72106A may not be on the pre-main sequence, it is certainly 
one of the youngest known main sequence stars of its type.

\begin{table}
\centering
\caption{Fundamental physical properties of HD 72106A and B. Magnetic field parameters are determined in Section \ref{Doppler Imaging}. }
\begin{tabular}{ccc}
\hline \hline \noalign{\smallskip}
                &  HD 72106A         &  HD 72106B  \\
\noalign{\smallskip} \hline \noalign{\smallskip}
\teff\,(K)      & $11000 \pm 1000$   & $8750 \pm 500$ \\
\lgg\, (cgs)    & $4.0 \pm 0.5$      & $4.0 \pm 0.5$ \\
$L$ ($L_\odot$) & $22\,^{+20}_{-10}$ & $9.2\,^{+8.2}_{-4.3}$ \\
$R$ ($R_\odot$) & $1.3 \pm 0.5$      & $1.3 \pm 0.5$  \\
$M$ ($M_\odot$) & $2.4 \pm 0.3$      & $1.9 \pm 0.2$  \\
age (Myr)       & 6 -- 13            & 6 -- 13        \\
\vs\, (\kms)    & $41.0 \pm 0.3$     & $53.9 \pm 1.0$ \\
$P_{rot}$ (days)& $0.63995 \pm 0.00009$ & -- \\
$i (\degr)$   & $24 \pm 10$  & -- \\
$\beta (\degr)$& $57 \pm 5 $      & -- \\ 
$B_{\rm p}$ (G) &  $1230 \pm 80 $    & -- \\ 
\noalign{\smallskip} \hline \hline
\end{tabular}
\label{physical prop tab}
\end{table}

The physical properties we derive for both stars are summarized in Table \ref{physical prop tab}. 
It is worth noting that, while the absolute luminosities of the components 
are poorly determined, their ratio ($L_A/L_B = 2.3 \pm 0.4$) is very well determined.  
This is because the major uncertainty in the absolute luminosity is the distance to the stars, 
and the stars are located at the same distance.  Thus the values of 
the radii or masses derived for the two stars are not independent,
since spacing between the components in $\log L$ on the H-R diagram must remain fixed.

\section{Least Squares Deconvolution and Stellar Rotation}
\label{Rotation and Magnetic Field Geometry}

\subsection{LSD and Longitudinal Magnetic Field}
In order to measure the magnetic fields of HD 72106A \& B, 
we employed Least Squares Deconvolution  \citep[LSD;][]{Donati1997-major}.  
This cross correlation technique uses a table of input atomic data (the line mask) to 
produce a deconvolved mean line profile from an observed spectrum.
Using a large number of lines,
$\sim$7000 for our spectra, can result in dramatic improvements in S/N.  
LSD was performed on all our observations of the individual components of HD 72106, 
as well as the reconstructed spectra of HD 72106A.  
The full set of LSD profiles in Stokes $I$ and $V$ for HD 72106A, calculated using specific elements, are presented in Section \ref{Doppler Imaging} 
in Figure \ref{DI lsd fits}. 

The line mask used for the primary was calculated assuming an effective temperature of 11000 K, \lgg\, of 4.0, 
typical Ap star abundances, and a line depth cutoff of 0.1 (as a fraction of the continuum).  
This produced LSD profiles with a mean wavelength of 523.4 nm, a mean Land\'e factor of 1.27, 
and a mean excitation potential of 5.97 eV, with small variations between observations.  
These mean values represent the average values of the lines used in the computation of the LSD profile, weighted by the S/N of the lines.
The computed LSD profiles are relatively insensitive to variations or errors in the line mask in 
temperature or abundance \citep[e.g][]{Shorlin2002}.
For the secondary, a \teff~=~8750~K, \lgg~=~4.0 line mask was used with solar abundances and a 
line depth cutoff of 0.1.  With this mask we found a mean wavelength of 533 nm, 
a mean Land\'e factor of 1.21, and a mean excitation potential of 3.99 eV. 

The longitudinal magnetic field was then measured from each set of LSD profiles, using
Eq.~(1) of \citet{Wade2000-highPrecision-correctBz}.  
Integration was performed through the portion of the line profile that 
exceeded 15\% of the total line depth, with an additional 5 \kms\, on either end.
The longitudinal field data are presented in Table \ref{table_obs}.  
We find a maximum longitudinal field for HD 72106A of $374 \pm 32$~G. 

In the secondary, in contrast, we detect no longitudinal field, with error bars of 50--150~G. 
Moreover, the LSD profiles of HD 72106B yield no evidence of a magnetic field.  
The analysis of HD 72106B by \citet{Wade2007-HAeBe_survey}, discussed in Section \ref{HD 72106 intro}, 
found no magnetic field in either the Balmer lines or metallic lines with error bars of $\sim$100~G
Thus we conclude that if the secondary has a magnetic field, its longitudinal component never exceeds 
$\sim 200$ G.

\subsection{Rotation Period of HD 72106A}
\label{Rotation Period of HD 72106A}

Variability in the spectrum of HD 72106A was noted by \citet{Wade2005-HAeBe_Discovery}.  
A careful reexamination of their data shows that most of their reported spectrum variability 
was due to the lack of atmospheric dispersion correction during the acquisition of one of their spectra.  
However, we do see substantial metal line variability in all of our spectra of the primary, 
with a variation timescale on the order of a day.  
We do not confirm the rotational period of approximately 2 days proposed by \citet{Wade2005-HAeBe_Discovery}. 

\begin{figure}
\centering
\includegraphics[width=3.3in]{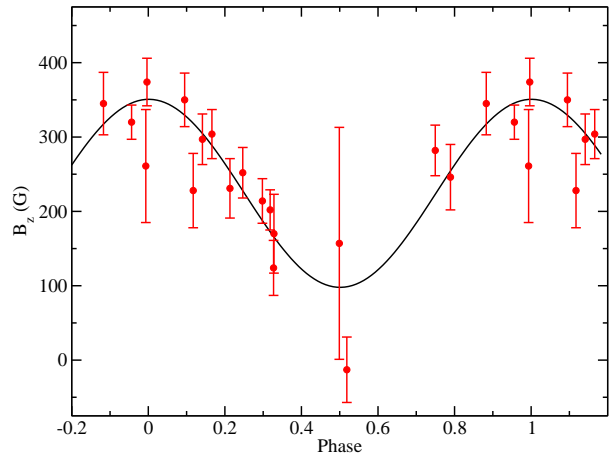} 
\caption{Longitudinal magnetic field measurements of HD 72106A (points), 
phased with the adopted 0.63995 day period, and 
the longitudinal field curve from the best fit dipolar magnetic geometry (solid line) determined in Section \ref{Doppler Imaging}. }
\label{phased Bz}
\end{figure}

\begin{figure*}
\centering
\includegraphics[width=3.3in]{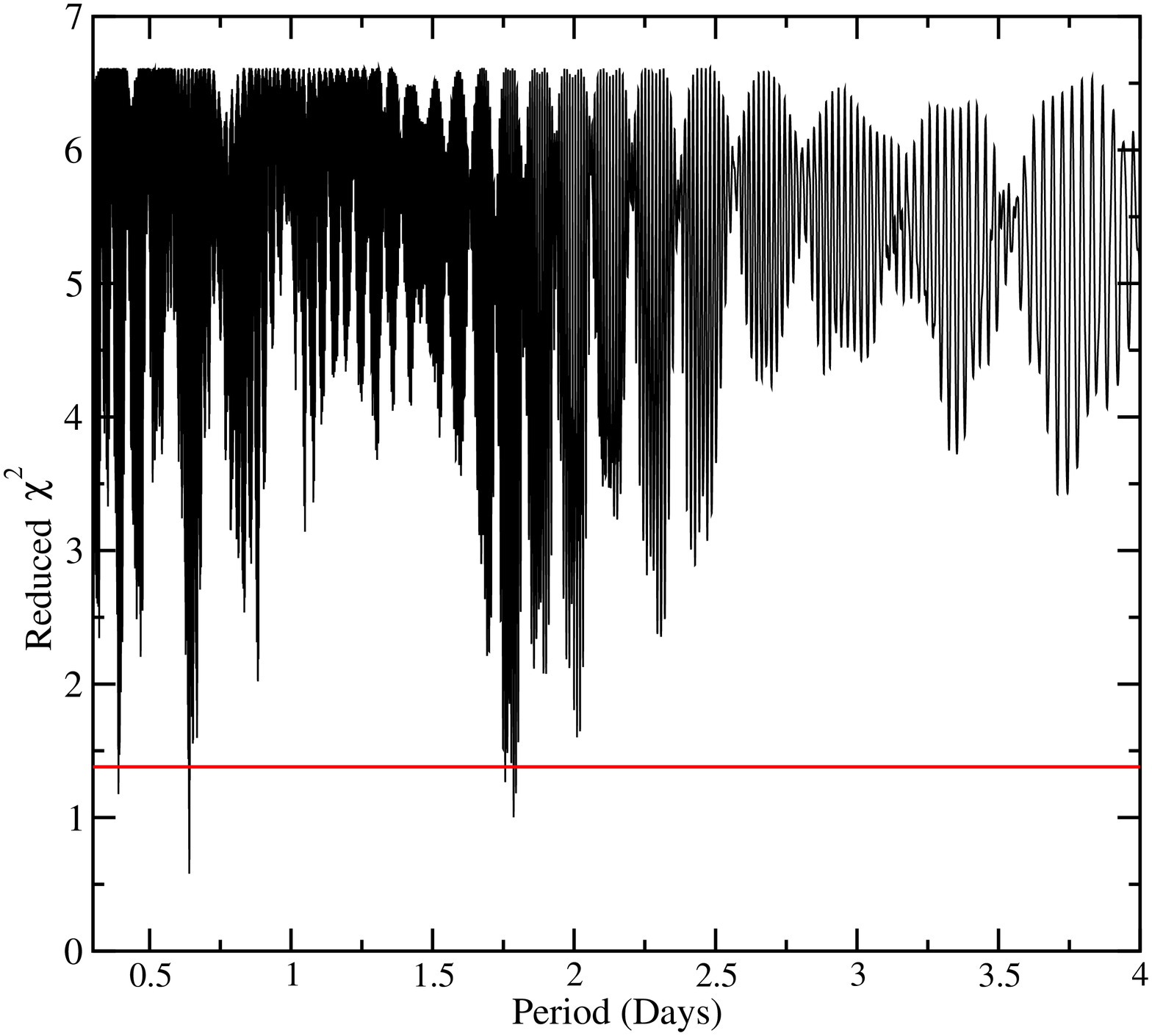} 
\includegraphics[width=3.3in]{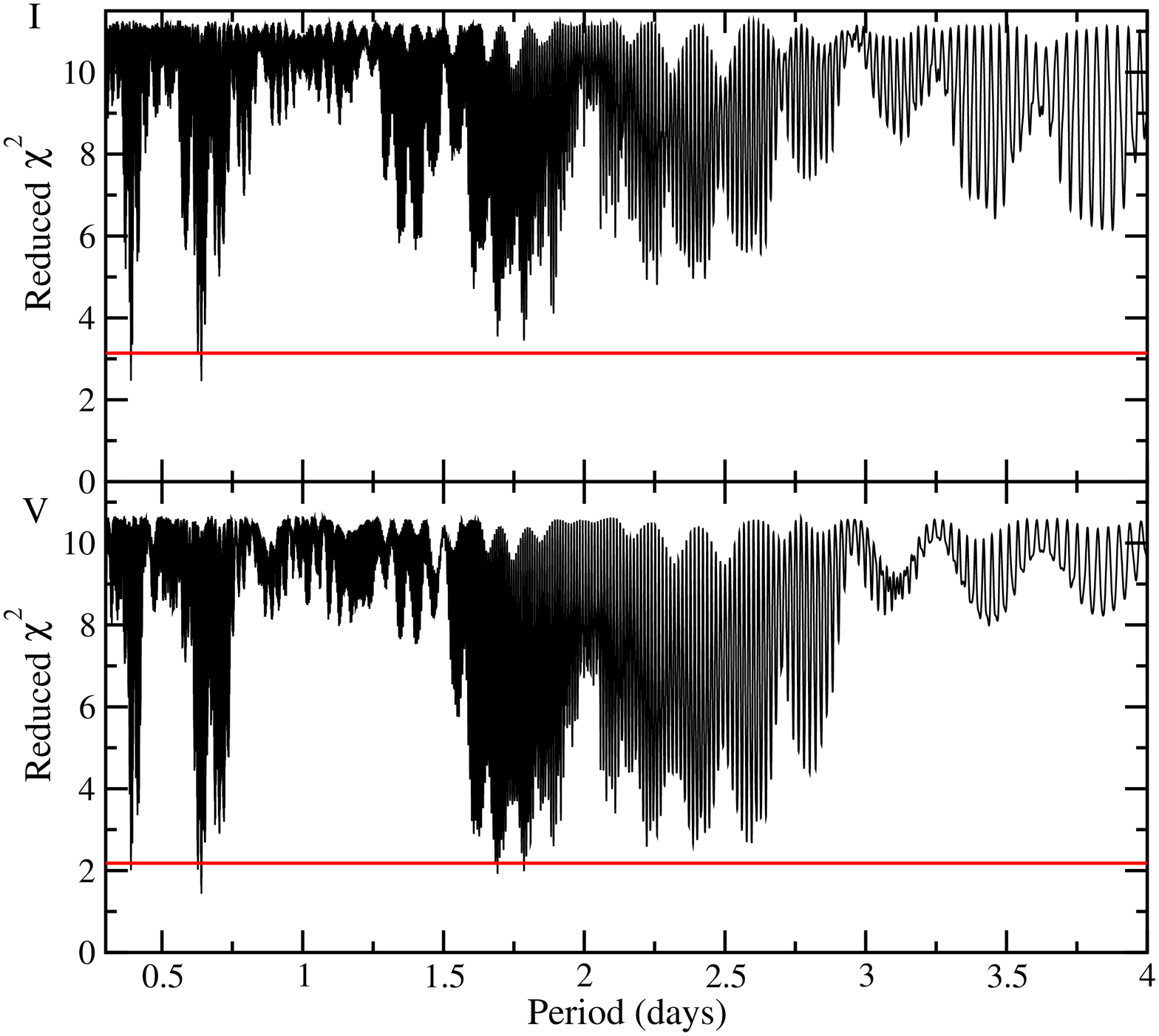}
\caption{ Periodograms for HD 72106A based on longitudinal field measurements (left panel) and 
LSD profile variability (right panel) in Stokes $I$ and $V$. A few minima in each periodogram fall 
below the 99\% confidence limit (horizontal line), however the deepest minimum is at 0.64 days in all cases. }
\label{periodograms}
\end{figure*}

In order to investigate whether this variability in HD 72106A could be due to rotational modulation, 
we fit the longitudinal magnetic field observations with a first order sinusoid of fixed period, 
but free phase, amplitude, and mean. 
This process was repeated for many periods to construct a periodogram in a fashion 
analogous to the Lomb-Scargle method \citep{numerical-recipes-Fortran}.  In the resulting periodogram,
shown in Figure \ref{periodograms} (left frame), 
we found a number of periods that provided 
a good fit to our observations, with the best period at about 0.640 days.  
Our longitudinal field observations, phased with this best fit period, are shown in Figure \ref{phased Bz}.
We interpret this variability as being due to the rotation of HD 72106A, 
in the context of the oblique rotator model \citep[ORM; e.g.][]{Stibbs1950-oblique_rot}.  
Using the radius determined in Section \ref{Fundimental Physical Properties}, 
together with the best fit period, we find that a range of inclination angles 
will satisfy the derived \vs\, from Section \ref{Primary's Abundances}. 
The smooth sinusoidal behavior of the longitudinal magnetic field, the stability of the period 
over two years, and the period consistent with the star's \vs\, and radius  
all support our interpretation of the variability in the context of the ORM. 
It is worth noting that the phase coverage for periods around 0.64 days is poor between phases 0.4 and 0.7.  
However, the remainder of our analysis is based on rotationally broadened LSD profiles rather than 
longitudinal field measurements, which mitigates this phase gap.

In order to search further for periodic variations in the observations of HD 72106A, we examined 
the variability within the Stokes $I$ and $V$ LSD profiles.  To do this, we 
constructed a periodogram based on the variability of each pixel  
of the LSD profiles, using the same technique as for the longitudinal field measurements.  
The periodograms for each pixel were then averaged and 
weighted by the amplitude of variability in that pixel, to produce a 
periodogram for the entire set of LSD profiles.  This procedure was performed for both 
Stokes $I$ and Stokes $V$ LSD profiles, and the resulting periodograms are shown in Figure \ref{periodograms} (right frame). 

Periods at 0.38983 days, 0.63995 days, 1.6921 days and 1.7859 days 
all produced minima in one or more of the periodograms that were within 
the 99\% confidence limit of the global minimum.  
These candidate periods were each investigated by eye.  
Each period was used to determine the phasing of the LSD profiles, 
and the profiles were examined for variability physically consistent with rotation.  
For a physically sensible rotational variation, 
features must move smoothly across the profile from blue to red, and there must not be large changes in the 
profile with small changes in phase.  
Upon careful examination, the period of 0.63995 days was 
the only period to produce a physically sensible phasing of the $I$ and $V$ profiles. 
Thus we adopt the rotational ephemeris (with maximum longitudinal field at zero phase) for HD 72106A:
\begin{equation} 
{\rm HJD} = (2453747.017 \pm 0.013) + (0.63995\pm 0.00009) \cdot E.
\end{equation}
The LSD profiles phased according to this ephemeris are shown in Figure \ref{DI lsd fits}.
This period is notable in that it is one of the shortest rotation periods seen in any 
magnetic intermediate-mass star.

\section{Spectrum Synthesis}
\label{Spectrum Synthesis}

The ZEEMAN2 spectrum synthesis code was used to model the observed 
spectra of HD 72106A \& B.  This code solves the polarized radiative transfer 
equations in local thermodynamic equilibrium.  In order to find the best fit model of 
the observed spectrum, a Levenberg-Marquardt $\chi^2$ minimization routine was implemented.  
This routine consistently produced high quality, stable solutions with good efficiency.  
The results were, in all cases, carefully examined by eye to verify that they 
represented both a physical solution and the true best fit, rather than a local minimum in $\chi^2$.

Chemical abundances (in the form $\left[\frac{N_{\rm X}}{N_{\rm tot}}\right]$, assuming ${N_{He}}/{N_{H}} = 0.098$), 
projected rotational velocity (\vs), and for HD 72106B, microturbulence, 
were the free parameters in the fit.  Chemical abundances were assumed to be uniform 
both vertically and horizontally.  This is not strictly true for HD 72106A, which displays 
patchy distributions of some elements, discussed further in Section \ref{Doppler Imaging}.  
Thus the abundances derived for HD 72106A are averages across the visible disk of the 
star at the phase of the observation used (the 12 Jan. 06 observation, rotation phase 0.53).  
However, since there is not much change in equivalent widths between spectra, 
the observation of HD 72106A used is approximately representative of the 
global average for all elements, even those with patchy distributions.

Microturbulence is not expected in the primary, due to its strong magnetic field.  
By analogy to Ap/Bp stars, the magnetic field would likely suppress microturbulence, 
thus microturbulence was fixed at 0 \kms\, for HD 72106A.  This did not produce any 
detectable discrepancies between our model and observed spectra.
For the primary we adopted the magnetic field model derived from the Stokes $V$ LSD profiles 
in Section \ref{Doppler Imaging}.  
Uncertainties in this magnetic field model have a negligible impact on the derived abundances. 
For the secondary no magnetic field was included.   

Standard model atmospheres with solar abundance, computed with the ATLAS9 \citep{Kurucz1993-ATLAS9etc} 
code, were used as input for ZEEMAN2.  
Input atomic line data was obtained from the Vienna Atomic Line Database (VALD) \citep{Kupka1999-VALD}, 
using the ``extract stellar'' utility in VALD (with the default line list configuration).  

Seven segments of spectrum, ranging from 100 to 200~\AA\, in length, were independently fit 
for each star (4170-4265~\AA, 4400-4500~\AA, 4500-4600~\AA, 4600-4700~\AA, 5000-5200~\AA, 5200-5400~\AA, and 5400-5600~\AA) 
The final best fit parameters were taken as the 
average of the parameters derived from individual windows.  Uncertainties were 
taken as the standard deviation of the parameters.  For abundances that were only 
constrained in a few windows,  the uncertainties were estimated by eye, 
by taking the change in abundance necessary to shift the synthetic spectrum beyond the 
noise in the observed spectrum and any possible continuum normalization errors.  
In these case the presented uncertainties are 
indicated by an asterisk (*) in Table \ref{total abun tab}.

\subsection{Abundances in HD 72106A}
\label{Primary's Abundances}
\begin{figure*}
\centering
\includegraphics[width=3.4in]{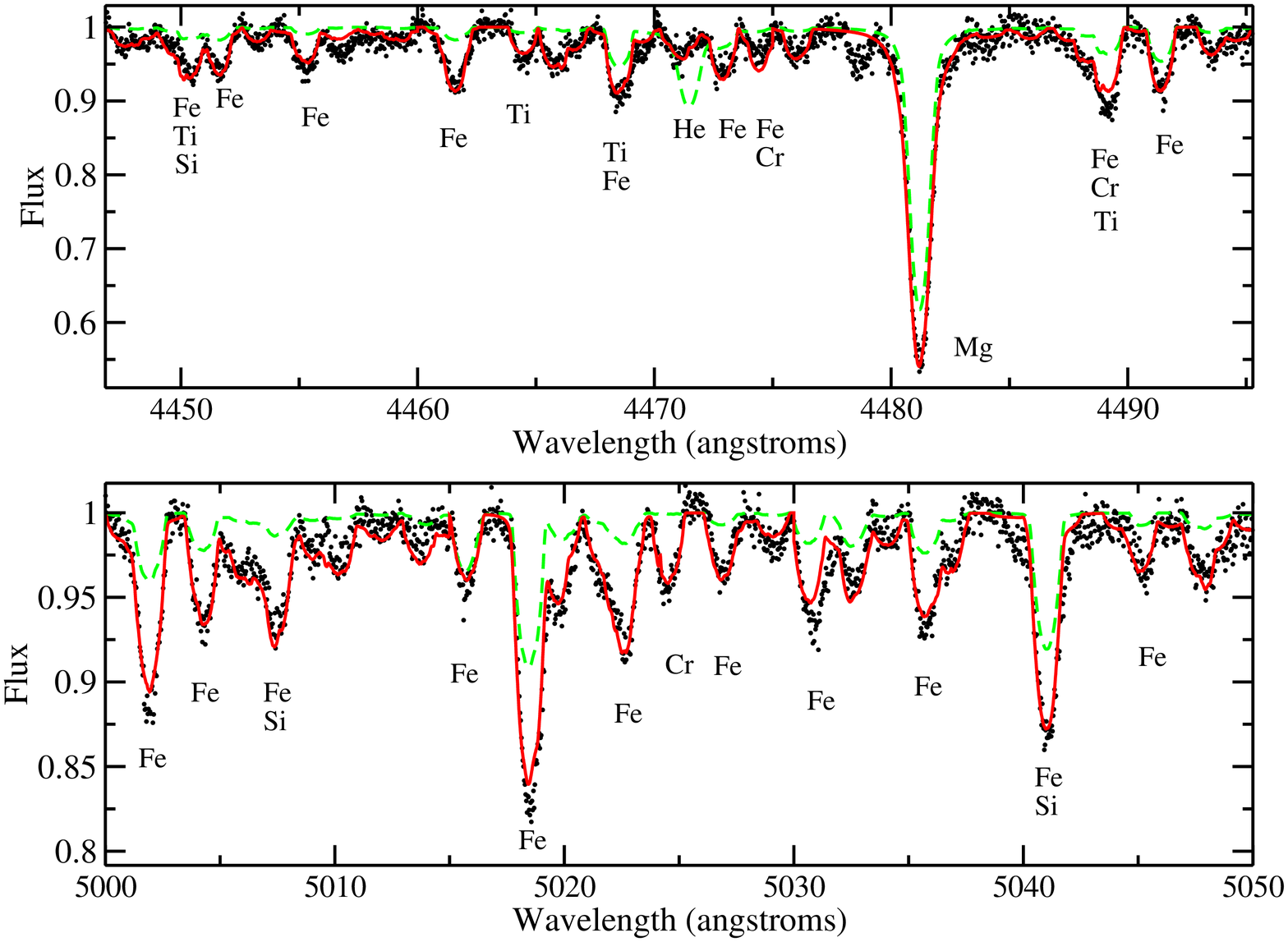}
\includegraphics[width=3.4in]{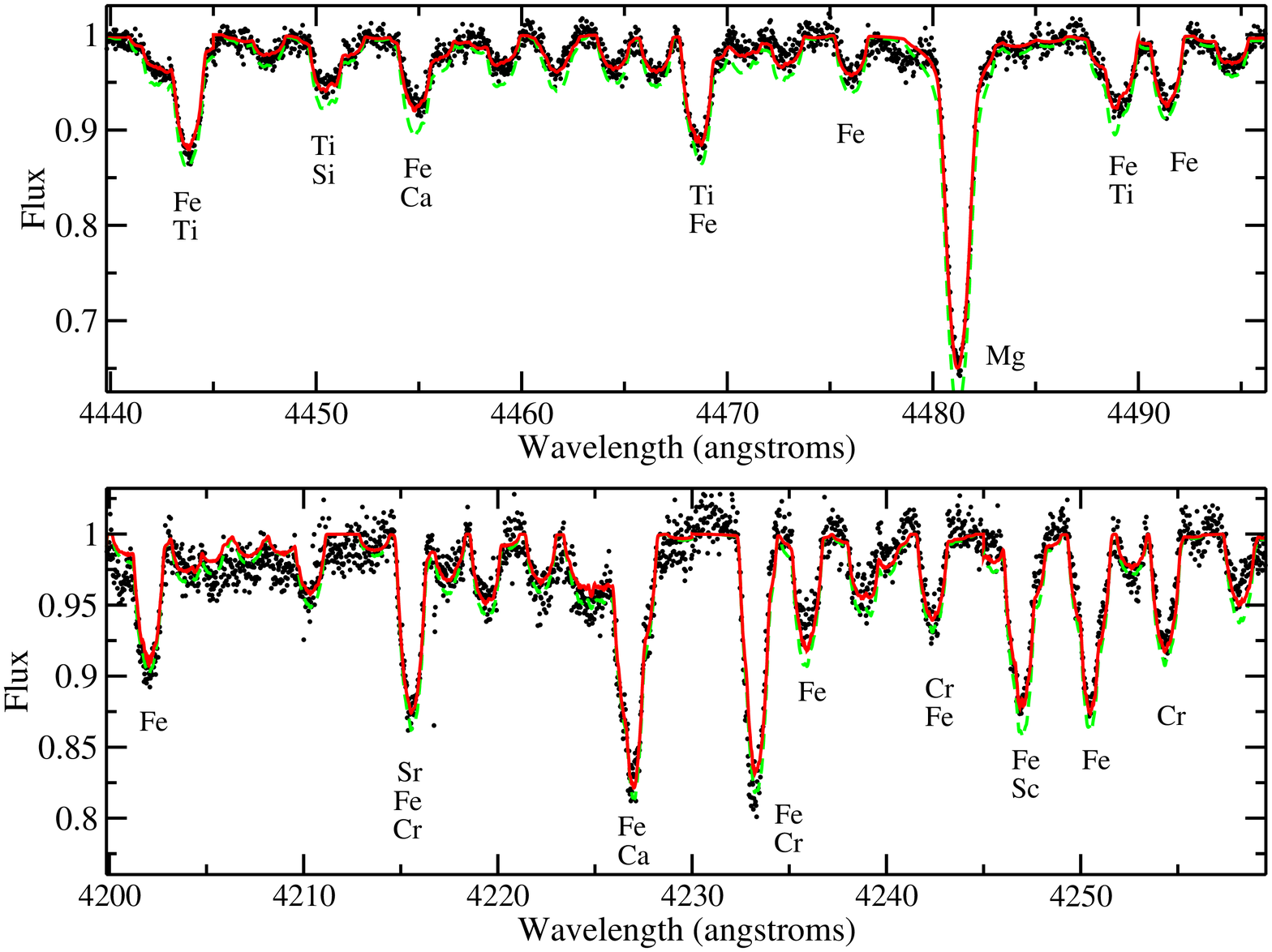}

\caption{ Sample best fit synthetic spectra for 
HD 72106A (left panels) and HD 72106B (right panels) in two independently fit spectral windows. 
The observations of the individual stellar components from Mar. 5, 2007 are used for both stars.
Major contributors to each line have been 
labeled, in order of importance.  The smooth 
solid line is the best fit spectrum in this region, the dashed line 
is a spectrum computed with solar chemical abundances. Dots represent the observations.  }
\label{spec fits}
\end{figure*}

Sample fits of our synthetic spectra to the observations of HD 72106A are presented in Figure \ref{spec fits}.  
The average best fit abundances and \vs, with uncertainties, 
are given in Table \ref{total abun tab}.  The abundances are presented graphically, 
relative to solar abundances, in Figure \ref{total abun plot}.  

Remarkably strong overabundances of Si, Cr, Fe and Nd are found in HD 72106A.  The Si and Fe abundances appear to 
be above solar by $\sim$1 dex, and Cr appears to be enhanced by $\sim$2 dex, whereas He appears to be 
$\sim$1.5 dex underabundant.  A number of elements, such as Al, Sc and Sr appear to
have solar abundances.  A couple of elements, particularly Mg and O, hint at 
possible peculiarity but require further study before concrete conclusions can be drawn.  
We find \vs\, = $41.0 \pm 0.7$ \kms, which is fairly low for a main sequence B star, 
but within the normal range for Bp stars.  The strong overabundances in Si, Cr, Fe, and Nd, as 
well as the underabundance in He, are common features of cooler Bp stars \citep{Jaschek-behavior_abun}.

\subsection{Abundances in HD 72106B}
\label{Secondary's Abundances}
\begin{figure}
\centering
\includegraphics[width=3.3in]{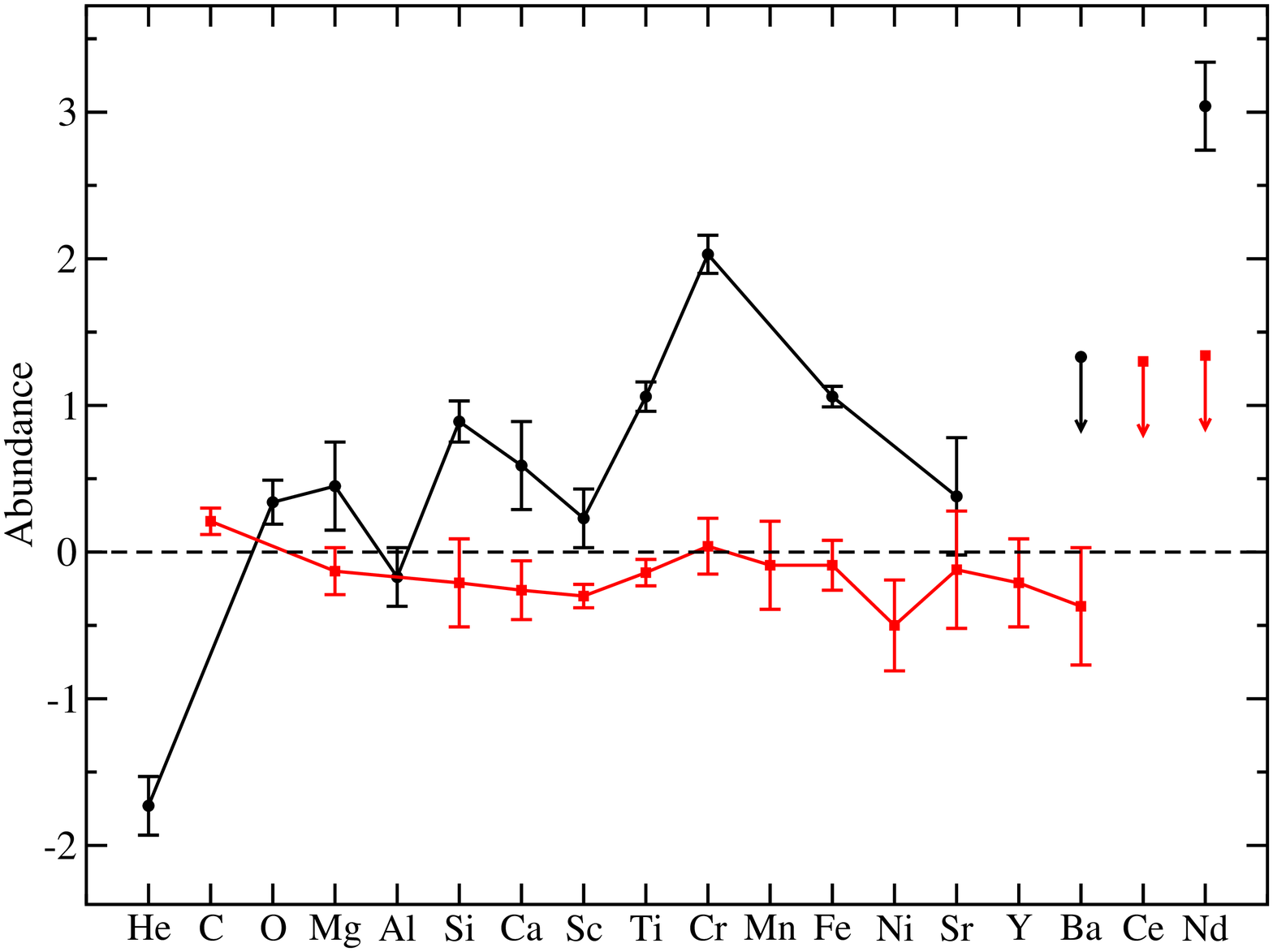}
\caption{Abundances relative to solar for HD 72106A 
(circles) and HD 72106B (squares), averaged over all spectral windows modeled.  
The dashed line at 0 represents solar abundance, based on \citet{Grevesse2005-solar_abun}.  
Points marked with an arrow indicate the value is an upper limit only.  
Strong departures from solar abundance can be seen for HD 72106A,
whereas HD 72106B has largely solar abundances. }
\label{total abun plot}
\end{figure}

HD 72106B was initially modeled with an effective temperature of 8000 K and \lgg\, = 4.5. 
However, as discussed in Section \ref{Fundimental Physical Properties}, 
this model provided a poor fit to the 
observed stellar spectrum, and hence \teff\, and \lgg\, were included as free 
parameters in our fit.  Chemical abundances, \vs\, and microturbulence were also 
included as free parameters in the fit, while the magnetic field was set to zero.
From this procedure we found a best fit \teff\, of $8750 \pm 500$ K and \lgg\, = $4.0 \pm 0.5$.  
Sample best fit synthetic spectra can be seen in Figure \ref{spec fits}, compared 
with the observed spectrum used in the fitting process.  The average best fit 
abundances, \vs, and microturbulence derived for HD 72106B are 
presented in Table \ref{total abun tab}, and the abundances are shown graphically 
relative to solar abundances in Figure \ref{total abun plot}.

The large majority of elements are consistent with solar abundances, 
within $2\sigma$ at most.  
A few elements appear to depart marginally from solar values, with a significance 
slightly greater than $2\sigma$.  C appears to be overabundant 
with $\sim 2\sigma$ significance, whereas Sc is underabundant by $\sim 3\sigma$.  
Under abundances of Sc and Ca are characteristic of Am stars. 
However, since the Ca abundance is approximately normal in HD 72106B, 
and iron peak elements are not overabundant, HD 72106B is not Am star. 
Thus, nearly all the elements are within $2\sigma$ of solar abundance, 
and no elements display the strong peculiarities seen in HD 72106A.  
This result is consistent with the approximately solar abundances seen in most HAeBe stars \citep{Acke2004-HAeBe_Abun}. 
It is of possible importance that the best fit abundances are consistently 
below solar, albeit with small significance levels.  However there 
remains a large uncertainty in our best fit temperature and \lgg. 
An increase in the adopted temperature by 250 K to 9000 K would 
systematically increase the best fit abundances by 0.1 dex or 0.2 dex, essentially eliminating this trend.

\begin{table}
\centering
\caption{Averaged best fit chemical abundances, \vs\, and microturbulence ($\xi$) for HD 72106A and B as 
well as solar abundances from \citet{Grevesse2005-solar_abun}. 
Entries marked by an * are based on only a few lines. }
\begin{tabular}{cccc}
\hline \hline \noalign{\smallskip}
         & HD 72106A      & HD 72106B            & Solar \\
\noalign{\smallskip} \hline \noalign{\smallskip}
\vs\, (\kms)& 41.0 $\pm$ 0.6&    53.9 $\pm$ 1.0  &        \\
$\xi$ (\kms)&              &     2.3 $\pm$ 0.6   &        \\
\noalign{\smallskip} \hline \noalign{\smallskip}
He      &  -2.8 $\pm$ 0.2 *&                     & -1.07  \\
C       &                  &  -3.40 $\pm$ 0.08   & -3.61  \\
O       &  -3.0 $\pm$ 0.15*&                     & -3.34  \\
Mg      &  -4.0 $\pm$ 0.3  &  -4.60 $\pm$ 0.16   & -4.47  \\
Al      &  -5.8 $\pm$ 0.2 *&                     & -5.63  \\
Si      & -3.60 $\pm$ 0.14 &  -4.7 $\pm$ 0.3     & -4.49  \\
Ca      &  -5.1 $\pm$ 0.3 *&  -6.0 $\pm$ 0.2     & -5.69  \\
Sc      &  -8.6 $\pm$ 0.2 *& -9.13 $\pm$ 0.08    & -8.83  \\
Ti      & -6.04 $\pm$ 0.10 & -7.24 $\pm$ 0.09    & -7.10  \\
Cr      & -4.33 $\pm$ 0.13 & -6.32 $\pm$ 0.19    & -6.36  \\
Mn      &                  &  -6.7 $\pm$ 0.3    *& -6.61  \\
Fe      & -3.49 $\pm$ 0.07 & -4.64 $\pm$ 0.17    & -4.55  \\
Ni      &                  &  -6.3 $\pm$ 0.3     & -5.77  \\
Sr      &  -8.7 $\pm$ 0.4 *&  -9.2 $\pm$ 0.4    *& -9.08  \\
Y       &                  & -10.0 $\pm$ 0.3     & -9.79  \\
Ba      & $\leq -8.5$     *& -10.2 $\pm$ 0.4    *& -9.83  \\
Ce      &                  &  $\leq$ -9.0       *& -10.30 \\
Nd      &  -7.5 $\pm$ 0.3 *&  $\leq$ -9.2       *& -10.54 \\

\noalign{\smallskip} \hline \hline
\end{tabular}
\label{total abun tab}
\end{table}

\section{Magnetic Doppler Imaging}
\label{Doppler Imaging}

The observed metallic line variability with rotation strongly suggests 
that there are horizontal chemical abundance inhomogeneities in the atmosphere of HD 72106A.  
In an effort to determine the structure of these inhomogeneities, as well as the magnetic field structure, we  performed 
Magnetic Doppler Imaging (MDI) on HD 72106A.  Doppler Imaging is a method for inverting a timeseries of variable line profiles at 
known rotational phases in order to reconstruct the surface distributions that give rise to the observed 
variations.  In MDI, polarized line profiles are also included in the inversion, 
allowing for the reconstruction of a magnetic field through the Zeeman effect. 

We used the MDI code INVERS10, developed by O. Kochukhov and 
N. Piskunov \citep{Piskunov2002-MDI-intro1,Kochukhov2002-MDI-intro2} for surface chemical 
abundance mapping and magnetic field reconstruction.  This program 
performs accurate polarized LTE spectrum synthesis, using pre-calculated model stellar 
atmospheres.  INVERS10 allows for simultaneous modeling of multiple 
chemical elements and multiple wavelength regions, and takes into account blended lines.  
Tikhonov regularization is used \citep{Tikhonov1963,Piskunov2002-MDI-intro1}, to constrain 
abundance gradients in the surface map.  While Tikhonov regularization can be applied to the magnetic 
field as well, we opted for multipolar regularization \citep{Kochukhov2002-MDI_alpha2CVn}, 
due to the lack of Stokes $Q$ and $U$ spectra and the relatively low S/N in the Stokes $V$ spectra. 

Input into the MDI model of HD 72106A included the adopted effective temperature and surface gravity 
from Section \ref{Fundimental Physical Properties}: \teff\, = 11000 K and \lgg\, = 4.0. 
The inclination angle of the star's rotation axis to the line of sight ($i$) was calculated using 
the radius determined in Section \ref{Fundimental Physical Properties}, the rotation period 
from Section \ref{Rotation Period of HD 72106A}, and the \vs\, determined from spectrum fitting 
in Section \ref{Primary's Abundances}, giving an angle of $i = 24 \pm 10\degr$.   
The phasing of observations was determined from the adopted ephemeris. 
An initial \vs\, of 41 \kms\, and abundances, for the treatment of blended lines, were 
taken from the results of our abundance analysis of HD 72106A, presented 
in Section \ref{Primary's Abundances}.  
The initial magnetic field geometry was assumed to be a dipole.

Initially, five Si II lines at 4128 \AA, 4130 \AA, 5056 \AA, 5041 \AA, and 6371 \AA\, 
were considered for Doppler Imaging.  However, the S/N of the observations was insufficient to 
produce high quality Doppler maps for Si, as well as for all other elements.  

In order to improve the S/N of our observations we calculated LSD profiles for individual chemical elements, 
and then performed MDI using those profiles.  
The use of LSD profiles to determine magnetic field geometries, stellar pulsation, 
and surface structures \citep{Donati2000-cool-lsd-spots-ex, Donati2001-betaCeph} is well established. 
LSD profiles of individual elements were 
obtained by creating LSD line masks containing only lines of the element of interest.  
The line masks used were derived from the 11000 K Ap star line mask discussed in Section \ref{Rotation and Magnetic Field Geometry}.
Doppler Imaging was then performed using the set of LSD profiles for each element as described above.  
The elements Si, Ti, Cr, and Fe were used in this process.  
Due to the low S/N in Stokes $V$ for the purposes of MDI, even in the LSD profiles, 
the magnetic field geometry was determined from the higher S/N Cr and Fe profiles, 
and held fixed for the lower S/N Si and Ti profiles.  Additionally, the 
multipolar magnetic field regularization was restricted to to $l = 1$ modes, effectively providing a `dipolar regularization'. 
Mean LSD profile atomic data were used for the Doppler Imaging process, 
providing wavelengths and excitation potentials, shown in Table \ref{LSD-DI atomic data}.  
Since the depth of the LSD profile is a complex function of the lines used in the analysis, 
the mean $gf$ value may not represent a realistic oscillator strength for the LSD profile. 
As a consequence, the absolute abundance scale of the Doppler images is somewhat uncertain. 

\begin{figure*}
\centering
\includegraphics[width=5.0in]{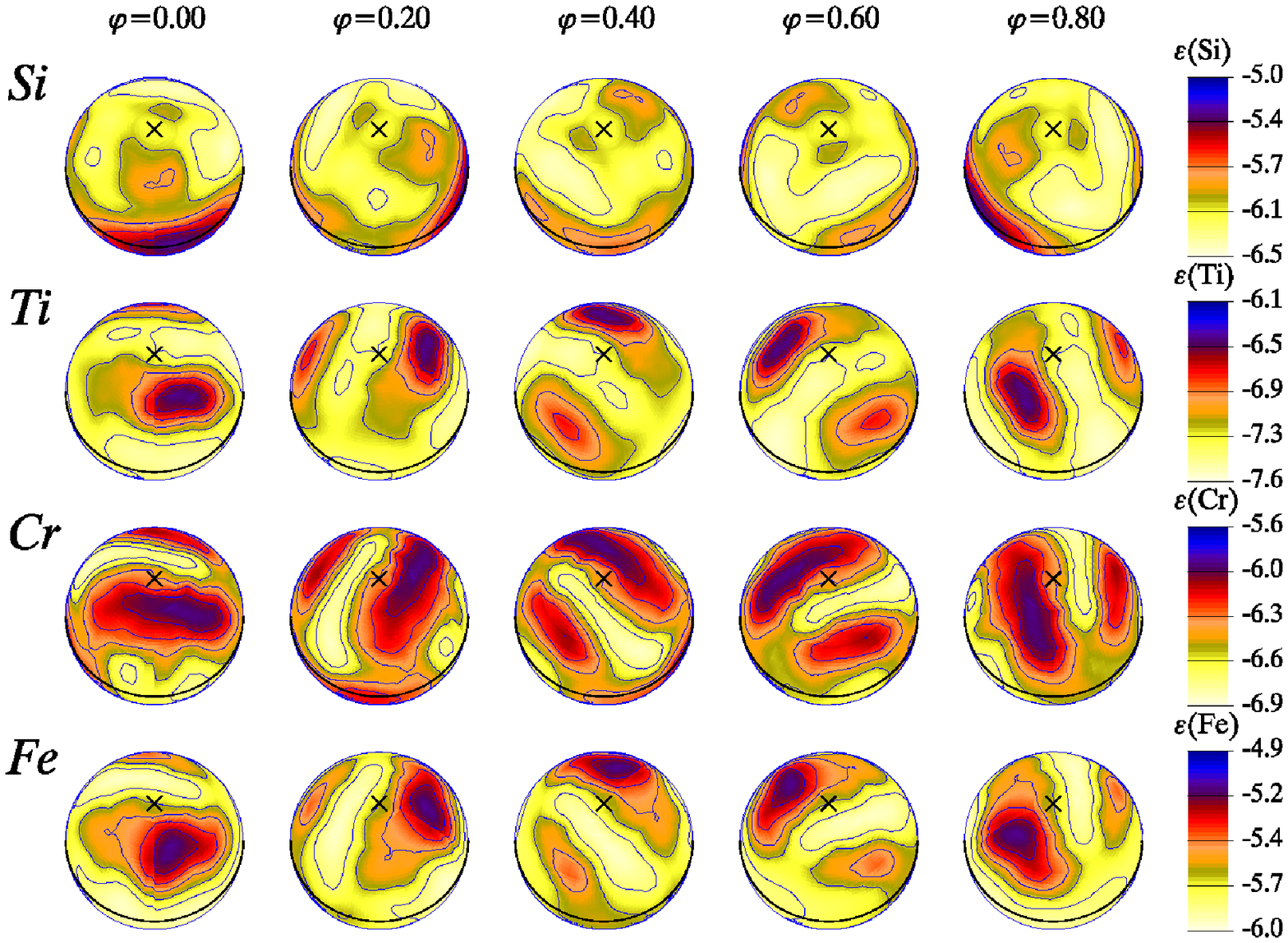}
\includegraphics[width=5.0in]{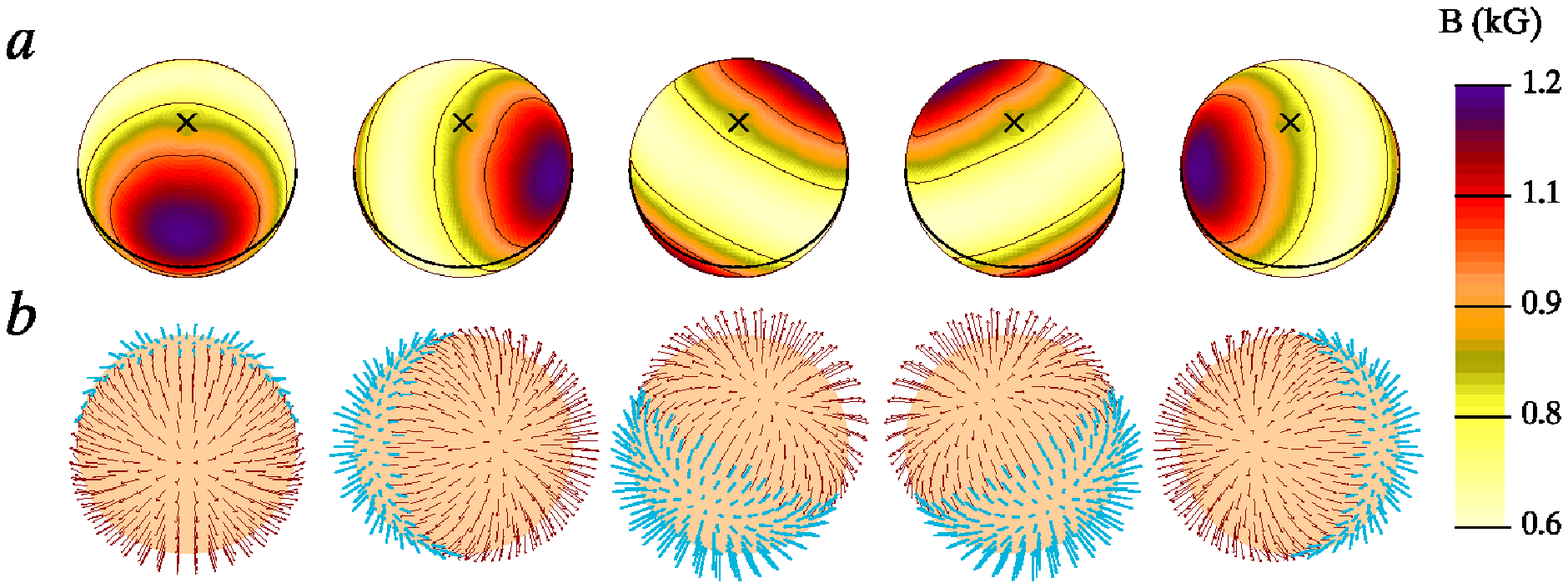}
\caption{Surface maps of Si, Ti, Cr, and Fe abundances, as well as the magnetic field, for HD 72106A.  
The maps are all based on fits to LSD profiles.   The `X' represents the 
rotational pole, the thick line circle indicates the rotational equator.  
The scale on the right in the abundance maps is in units of log $\frac{\rm N_{X}}{\rm N_{tot}}$.
The map of the magnetic field intensity is labeled `$a$' and the magnetic field direction is `$b$' (shown as arrows). } 
\label{DI lsd maps}
\end{figure*}

\begin{figure*}
\centering
\includegraphics[height=6.0in]{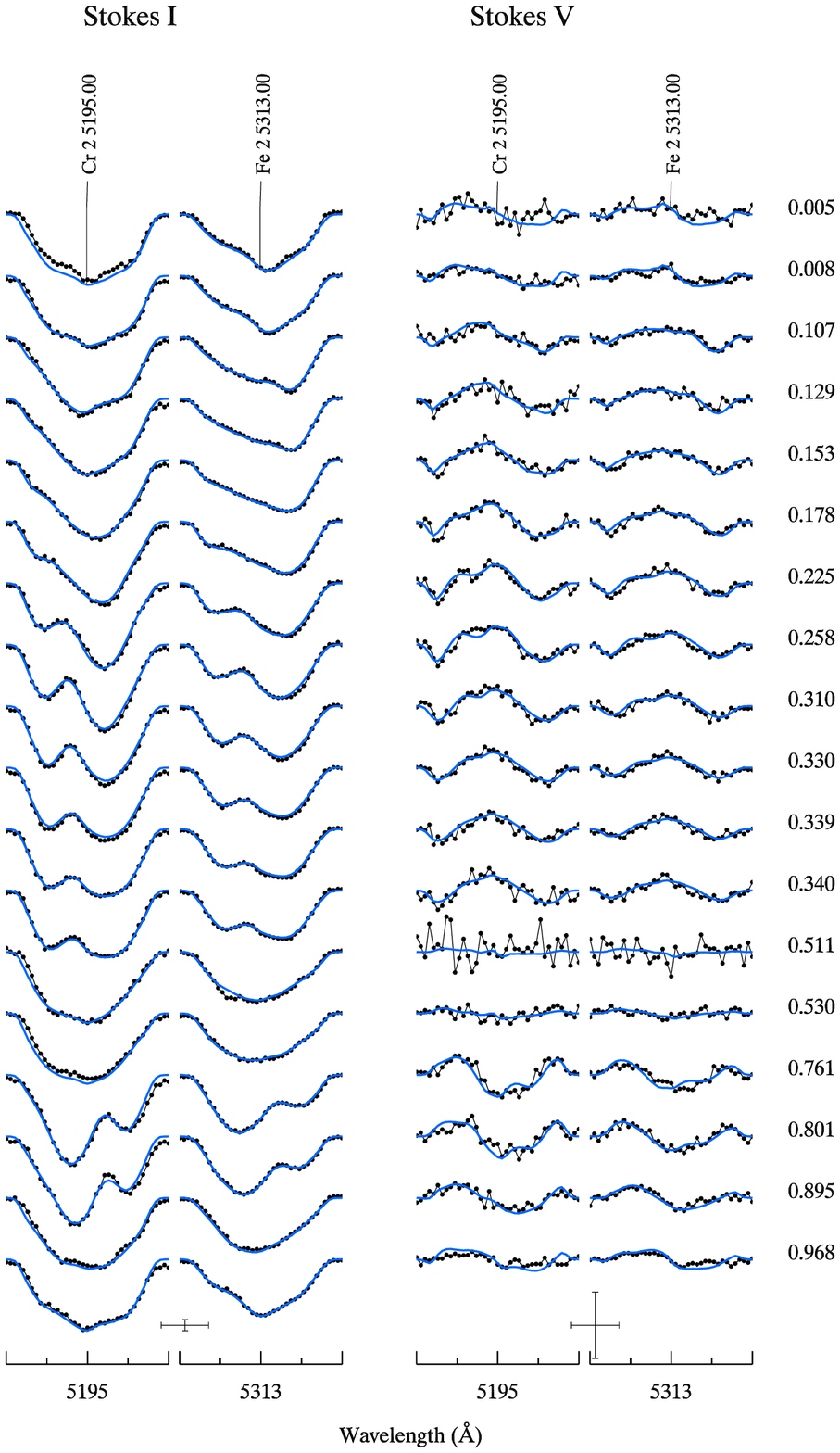}
\includegraphics[height=6.0in]{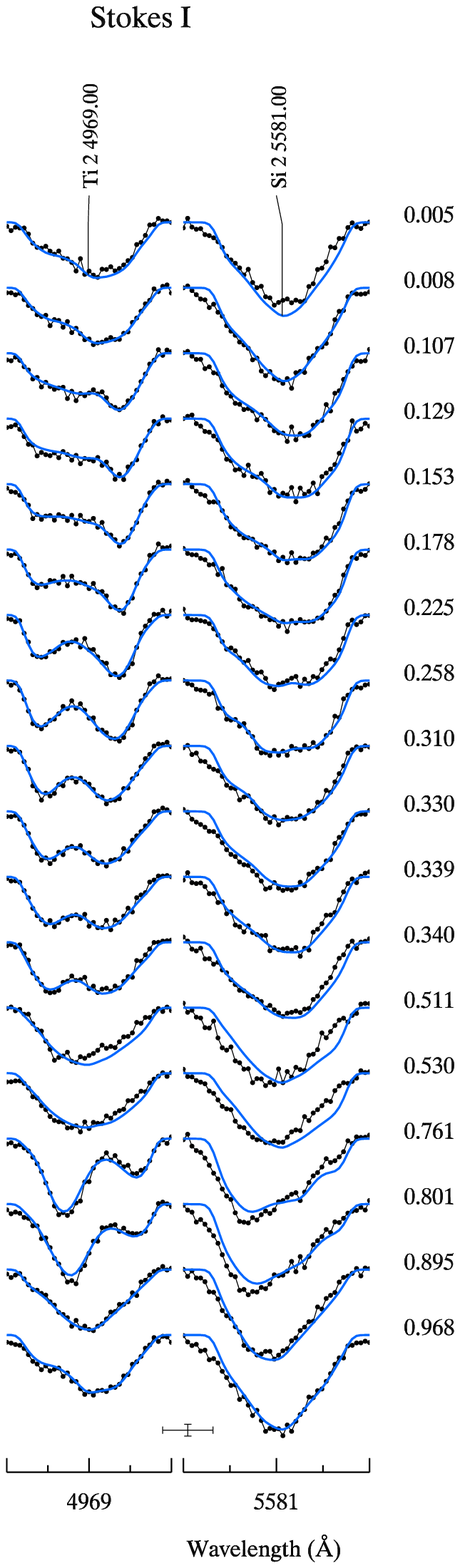}
\caption{Fits of synthetic Stokes $I$ (for Si, Ti, Cr, and Fe) and Stokes $V$ (for Cr and Fe) LSD profiles to observed LSD profiles. 
The profiles are labeled according to element, phase and mean wavelength.  The bars near the bottom of the figure 
indicate the vertical and horizontal scale, 0.5\% of the continuum and 0.5 \AA\, respectively.
Generally good fits can be seen, ranging from the high S/N Fe 
profiles at the best to the noisy Si profiles at the worst.  }
\label{DI lsd fits}
\end{figure*}

The final abundance maps, and magnetic modulus and vector maps, are shown in Figure \ref{DI lsd maps}.  
The corresponding best fits to the LSD line profiles are presented in Figure \ref{DI lsd fits}.   
In this process we have assumed that the LSD profile behaves like a real spectral line, 
and that the mean atomic data is approximately representative of the LSD profile. 
Additionally, unaccounted for line blends in the LSD process could slightly distort the LSD profiles, 
adding some uncertainty to the finer details of the Doppler map.  
Thus, while the large scale features of the Doppler Imaging process are fairly certain, one 
should use some caution in interpreting the finer details of the map, as well as the absolute abundance scale. 
Similarly, the use of LSD profiles to reconstruct the magnetic field geometry through MDI could potentially introduce
some systematic uncertainty into the derived geometry.  However, the magnetic geometry derived 
through MDI is fully consistent with the longitudinal magnetic field variability and 
nearly identical to that we derive by directly modeling the LSD Stokes V profiles with a simple dipole.  

\begin{table}
\centering
\caption{Mean atomic data for the element specific LSD profiles used in Magnetic Doppler Imaging. }
\begin{tabular}{ccccc}
\hline\hline
                               & Si    & Ti    & Cr    & Fe   \\
\hline
mean wavelength (\AA)          & 5580  & 4970  & 5190  & 5310 \\
mean excitation (eV)           & 7.68  & 2.03  & 5.65  & 6.46 \\
$\log$ mean $gf$               & -0.17 & -0.63 & -0.26 & -0.35\\
mean Land\'e factor            &  1.35 &  1.12 &  1.29 &  1.27 \\
\hline\hline
\end{tabular}
\label{LSD-DI atomic data}
\end{table}

Strong inhomogeneities are reconstructed for all four elements.  
Ti, Cr, and Fe all seem to share similar abundance patterns, 
though Cr appears to have somewhat larger, more elongates spots.  
The similar distributions are reflected in the similar phase variations of the LSD profiles 
in Figure \ref{DI lsd fits}.
A large patch of overabundance centered near 
phase 0 is apparent in all three maps, with another somewhat 
smaller overabundance spot about $180\degr$ away in 
longitude, at the same latitude, around phase 0.6.  
The Si map shares the larger spot but not the smaller, and this is 
reflected in the Si LSD profiles.  
There appears to be a large overabundance spot of Si at the equator (seen near the limb) near phase 0, although the 
sensitivity of the map is poor in that region due to its small projection. 
Note that the fits to the relatively noisy Si LSD profiles are somewhat poorer than the fits for the other elements analyzed,
thus the results for Si are somewhat more uncertain. 
The magnetic field geometry derived, shown in Figure \ref{DI lsd maps}, 
is essentially described by a dipole with a strength at the magnetic pole ($B_{\rm p}$) of $1230 \pm 80$~G, 
and an obliquity angle of the magnetic dipole with respect to the rotation axis ($\beta$) of $57\degr \pm 5$. 
This dipole magnetic field geometry is in full agreement with the observed longitudinal magnetic field variability.  
If there are any departures from a purely dipolar field geometry, they are not evident in our data.  
When the magnetic field geometry is compared to the abundance maps, it appears that the 
positive magnetic pole lies near the large spot of 
overabundance at phase 0 in all four maps.  
However, the magnetic pole is offset from the center of the abundance spot, 
being nearer the rotational equator and at a slightly different latitude, 
thus the relationship is not entirely clear. 

Interestingly, the pattern of abundances seen here for HD 72106A bear some similarities 
to those obtained for $\epsilon$ Ursae Majoris by \citet{Lueftinger2003-DI_epUMa}.
$\epsilon$ Ursae Majoris (HD 112185) is a 9000 K (\lgg\, = 3.6) late main sequence Ap star 
with $\sim 5$ day period and a dipole field strength of several hundred gauss.  
\citet{Lueftinger2003-DI_epUMa} constructed Doppler maps of Ti, Cr and Fe, as well as Ca, Mg, Sr, and Mn. 
They found distributions of Cr and Fe very similar to each other, with 
two large spots of overabundance near the longitude of the magnetic poles. 
Ti was roughly anti-correlated with Fe and Cr, displaying two large spots of 
underabundance at the same positions as the overabundance spots of Cr and Fe.   
We see little variability in Ca, and have insufficient S/N in Ca, Mg, Sr, and Mn to 
construct Doppler maps for comparison with \citet{Lueftinger2003-DI_epUMa}. 
The results for Fe and Cr are qualitatively very similar to what we see for HD 72106A, 
however the results for Ti differ significantly.  
The significance of these similarities and differences is unclear, 
as the evolution of chemical abundance spots over a star's main sequence lifetime is not understood.

\section{Discussion and Conclusions}

We have analyzed 20 high resolution spectropolarimetric observations 
of the HD 72106 system.  In these observations we see clear evidence 
for a magnetic field in HD 72106A.  We also confirm that HD 72106B 
is a HAeBe star, based on emission in H$\alpha$ and the O~{\sc i}~7771~\AA\, triplet, 
and that it displays no magnetic field.

There is strong evidence that the HD 72106 system is a true binary system.  
Both stars have the same proper motions, the same radial velocities,
and Hipparcos solution to the system places both stars at the same distance.
Although the separation of the components has remained constant in the pas 90 years, there 
has been a slow systematic increase in the position angle observed.
Additionally there is a wide range of possible bound orbits consistent with all available observations.  

We find the age of the system to be between 6 and 13 Myr, based on the H-R diagram position of the secondary 
(and assuming that, as a HAeBe star, the secondary is a pre-main sequence star).  
Thus the system is fairly evolved, for a pre-main sequence system.  
In the youngest limit, the primary would within 1 Myr of the ZAMS. 
However, in the oldest limit the primary would have reached the main sequence 9 Myr ago, 
but it would have only passed about 1.5\% of its main sequence lifetime.  Thus, while it may not 
be on the pre-main sequence, HD 72106A is certainly very young.  
While it is possible that the system is not coeval and instead was form by capture, 
the data available (consistent H-R diagram positions and the young age of the secondary) suggests that the system truly coeval. 
Even with the large uncertainty in age, only a few known Ap/Bp 
stars approach the maximum fractional age of HD 72106A 
\citep{Bagnulo2004-B_NGC2244-334, Landstreet2007-surveyFORS1}.
Given its evolutionary status, HD 72106A appears to represent a link between magnetic HAeBe stars and the Ap/Bp stars.  

HD 72106A possesses a strong, predominantly dipolar, magnetic field.  We find that a centered dipole 
with a polar field strength of $1230 \pm 80$ G and an obliquity angle of $57 \pm 5\degr$ is 
sufficient to provide a good fit to our observations.  In HD 72106B we find no evidence of a magnetic field,
with an upper limit on the longitudinal field of about 200 G.  
HD 72106A is one of the very youngest stars for which detailed modeling of the 
magnetic field geometry has been performed.  The only clearly younger object is the Herbig Be star HD 200775A, 
which \citet{Alecian2008-HD200775} find to have an age of $0.1 \pm 0.05$ Myr.
This star also displays a dipolar magnetic field geometry, 
with a magnetic field strength at the pole of $1000 \pm 150$ G and
an obliquity of $125 \pm 8\degr$, slightly offset 
at $0.05 \pm 0.04\,R_*$ from the star's center \citep{Alecian2008-HD200775}. 
HD 200775A and HD 72106A have similar magnetic field characteristics, 
both of which are analogous to most main sequence Ap/Bp stars.  
This provides further evidence that 
there is a continuum of magnetic A and B stars from the 
pre-main sequence through the main sequence.  

We see strong chemical peculiarities in HD 72106A, particularly an underabundance of He 
and overabundances of Si, Ti, Cr, Fe, and Nd.  Thus, from the point of view of abundance analysis,
HD 72106A appears to be a Bp star.  This implies that chemical peculiarities can form very early 
in a star's lifetime, possibly even on the pre-main sequence.  
This conclusion is supported by the recent results of \citet{Alecian2008-ClusterHAeBeLetter} 
who report chemical peculiarities in the spectrum of the HAeBe star W601 in NGC 6611.

The only other star with a comparably young fractional age in which strong chemical 
peculiarities have been measured in detail is the main sequence Bp star NGC 2244 334 \citep{Bagnulo2004-B_NGC2244-334}.  
This star has been on the main sequence for $2.3 \pm 0.3$ Myr, 
but is more massive and hence probably more evolved than HD 72106A, 
with a fractional age $\tau = 0.02 \pm 0.01$ \citep{Hensberge2000-Age_of_NGC2244,Bagnulo2004-B_NGC2244-334}.  
\citet{Bagnulo2004-B_NGC2244-334} find that the star has a temperature of $15000 \pm 1000$ K and 
an observed longitudinal magnetic field strength of 9000 G.  
The authors find a strong underabundance of He by $\sim1$ dex, 
as well as strong overabundances of Si and Fe by $\sim1$ dex, and Ti and Cr by $\sim2$ dex.  
These peculiarities are similar to those of HD 72106A, suggesting that there are important 
similarities between these stars.  

We find clear evidence for surface abundance inhomogeneities in HD 72106A. Magnetic Doppler Imaging 
suggests that there is a large spot of overabundance in Si, Ti, Cr, and Fe near the positive magnetic pole.  
These inhomogeneities are similar to those seen in Ap/Bp stars, 
further supporting a link between magnetic HAeBe stars and Ap/Bp stars.  
The Doppler reconstructions presented here are the earliest stage of
intermediate-mass stellar evolution ever mapped.

HD 72106B, in contrast to HD 72106A, appears to be chemically normal.  
While this is not the first binary system containing a magnetic chemically peculiar and 
a chemically normal star \citep[e.g.][]{Carrier2002-binarity_in_Ap},
it does raise the question of how such systems are produced.  Being a young binary, it is likely that the 
stars of HD 72106 formed at the same time from approximately the same material.  
The temperatures of these stars differ by 
only $2250 \pm 1120$ K, and the \vs\, values of the stars differ by only $\sim10$ \kms.  However, one star is 
magnetic and chemically peculiar while the other is not.  This suggests that, whatever mechanism 
gave rise to the difference between the stars, it must be rather sensitive to the particulars of 
the star's initial conditions.  

It is instructive to compare the characteristics of HD 72106A to those of 
the magnetic HAeBe stars HD 104237 and HD 190073.
HD 104237 was observed to possess a magnetic 
field by \citet{Donati1997-major} and confirmed by \citet{Donati2000-misc} and \citet{Alecian2008-CpAp_workshop}, 
with a longitudinal field strength of $\sim 50$ G (Donati, private communication). 
HD 104237 has a mass of about 2.3 $M_{\odot}$ and an age of about 2 Myr \citep{van_den_Ancker1998-HAeBe-photo/atmo}.
\citet{Acke2004-HAeBe_Abun} performed an abundance analysis of HD 104237 using equivalent widths, and found 
approximately solar abundances for a range of elements, including Si, Cr, and Fe.  
The star HD 190073 was reported to possess 
a magnetic field by \citet{Catala2007-HD190073}, with a longitudinal magnetic 
field strength of $74 \pm 10$ G.  \citet{Catala2007-HD190073} derive a mass of $2.85 \pm 0.25$ \Msun\, 
and an age of $1.2 \pm 0.6$ Myr (measured from the birth line) for this star.  
\citet{Acke2004-HAeBe_Abun} also studied the surface chemistry of this star and found roughly solar abundances. 

Thus it appears that the majority of known magnetic HAeBe stars are chemically normal, 
though some peculiar stars seem to exist \citep[such as NGC 6611 W601,][]{Alecian2008-ClusterHAeBeLetter}. 
This is in contrast to main sequence Ap/Bp stars, 
in which magnetic fields are nearly always found with chemical peculiarities.  
More analysis of chemical abundances in magnetic HAeBe stars must be performed, 
with an eye to identifying chemically peculiar objects. 
Interestingly, the young chemically peculiar stars HD 72106A and NGC 2244 334 display no emission, 
while HD 104237 and HD 190073 both display significant emission in their spectra.  
This suggests that HD 104237 and HD 190073 may still be undergoing significant accretion or mass loss 
while HD 72106A and NGC 2244 334 are not.  
Thus it may be that accretion or mass loss mixes the stellar atmosphere, 
inhibiting the buildup of chemical peculiarities through diffusion. 
However, once accretion halts chemical peculiarities may arise quickly.

\section*{Acknowledgments} 
Thanks to Brian Mason at the United States Naval Observatory for providing the
Washington Double Star Catalogue data on HD 72106, as well as commentary on that data.
CPF and GAW acknowledge support from the Academic Research Programme of the 
Canadian Department of National Defence. CPF, GAW, DH and JDL acknowledge 
Discovery Grant support from the Natural Sciences and Engineering Research Council of Canada.
EA is supported by the Marie Curie FP6 program.

\bibliography{masivebib.bib}{}

\begin{thebibliography}{}

\bibitem[\protect\citeauthoryear{{Acke} \& {Waelkens}}{{Acke} \&
  {Waelkens}}{2004}]{Acke2004-HAeBe_Abun}
{Acke} B.,  {Waelkens} C.,  2004, A\&A, 427, 1009

\bibitem[\protect\citeauthoryear{{Alecian}, {Catala}, {Wade}, {Donati},
  {Petit}, {Landstreet}, {B{\"o}hm}, {Bouret}, {Bagnulo}, {Folsom}, {Grunhut}
  \& {Silvester}}{{Alecian} et~al.}{2008a}]{Alecian2008-HD200775}
{Alecian} E.,  {Catala} C.,  {Wade} G.~A.,  {Donati} J.-F.,  {Petit} P.,
  {Landstreet} J.~D.,  {B{\"o}hm} T.,  {Bouret} J.-C.,  {Bagnulo} S.,  {Folsom}
  C.,  {Grunhut} J.,    {Silvester} J.,  2008a, MNRAS, 385, 391

\bibitem[\protect\citeauthoryear{{Alecian}, {Wade}, {Catala}, {Bagnulo},
  {Boehm}, {Bohlender}, {Bouret}, {Donati}, {Folsom}, {Grunhut} \&
  {Landstreet}}{{Alecian} et~al.}{2008b}]{Alecian2008-ClusterHAeBeLetter}
{Alecian} E.,  {Wade} G.~A.,  {Catala} C.,  {Bagnulo} S.,  {Boehm} T.,
  {Bohlender} D.,  {Bouret} J.-C.,  {Donati} J.-F.,  {Folsom} C.~P.,  {Grunhut}
  J.,    {Landstreet} J.~D.,  2008b, A\&A, 481, L99

\bibitem[\protect\citeauthoryear{{Alecian}, {Wade}, {Catala}, {Folsom},
  {Grunhut}, {Donati}, {Petit}, {Bagnulo}, {Marsden}, {Ramirez}, {Landstreet},
  {Boehm}, {Bouret} \& {Silvester}}{{Alecian}
  et~al.}{2008c}]{Alecian2008-CpAp_workshop}
{Alecian} E.,  {Wade} G.~A.,  {Catala} C.,  {Folsom} C.,  {Grunhut} J.,
  {Donati} J.-F.,  {Petit} P.,  {Bagnulo} S.,  {Marsden} S.~C.,  {Ramirez} J.,
  {Landstreet} J.~D.,  {Boehm} T.,  {Bouret} J.-C.,    {Silvester} J.,  2008c,
  in Proceedings of the CP\#AP Workshop Vol.~38 of Contributions of the
  Astronomical Observatory Skalnate Pleso, {Magnetism in pre-MS
  intermediate-mass stars and the fossil field hypothesis}.
pp 235--244

\bibitem[\protect\citeauthoryear{{Alecian} \& {Stift}}{{Alecian} \&
  {Stift}}{2007}]{AlecianG2007-el-distributions-atmos}
{Alecian} G.,  {Stift} M.~J.,  2007, A\&A, 475, 659

\bibitem[\protect\citeauthoryear{{Bagnulo}, {Hensberge}, {Landstreet},
  {Szeifert} \& {Wade}}{{Bagnulo} et~al.}{2004}]{Bagnulo2004-B_NGC2244-334}
{Bagnulo} S.,  {Hensberge} H.,  {Landstreet} J.~D.,  {Szeifert} T.,    {Wade}
  G.~A.,  2004, A\&A, 416, 1149

\bibitem[\protect\citeauthoryear{{Carrier}, {North}, {Udry} \&
  {Babel}}{{Carrier} et~al.}{2002}]{Carrier2002-binarity_in_Ap}
{Carrier} F.,  {North} P.,  {Udry} S.,    {Babel} J.,  2002, A\&A, 394, 151

\bibitem[\protect\citeauthoryear{{Catala}, {Alecian}, {Donati}, {Wade},
  {Landstreet}, {B{\"o}hm}, {Bouret}, {Bagnulo}, {Folsom} \&
  {Silvester}}{{Catala} et~al.}{2007}]{Catala2007-HD190073}
{Catala} C.,  {Alecian} E.,  {Donati} J.-F.,  {Wade} G.~A.,  {Landstreet}
  J.~D.,  {B{\"o}hm} T.,  {Bouret} J.-C.,  {Bagnulo} S.,  {Folsom} C.,
  {Silvester} J.,  2007, A\&A, 462, 293

\bibitem[\protect\citeauthoryear{Donati}{Donati}{2000}]{Donati2000-misc}
Donati J.-F.,  2000, Th\`ese d'habilitation, Observatoire Midi-Pyr\'en\'ees

\bibitem[\protect\citeauthoryear{{Donati}, {Mengel}, {Carter}, {Marsden},
  {Collier Cameron} \& {Wichmann}}{{Donati}
  et~al.}{2000}]{Donati2000-cool-lsd-spots-ex}
{Donati} J.-F.,  {Mengel} M.,  {Carter} B.~D.,  {Marsden} S.,  {Collier
  Cameron} A.,    {Wichmann} R.,  2000, MNRAS, 316, 699

\bibitem[\protect\citeauthoryear{{Donati}, {Semel}, {Carter}, {Rees} \&
  {Collier Cameron}}{{Donati} et~al.}{1997}]{Donati1997-major}
{Donati} J.-F.,  {Semel} M.,  {Carter} B.~D.,  {Rees} D.~E.,    {Collier
  Cameron} A.,  1997, MNRAS, 291, 658

\bibitem[\protect\citeauthoryear{{Donati}, {Semel} \& {Rees}}{{Donati}
  et~al.}{1992}]{Donati1992-ZeemanDI}
{Donati} J.-F.,  {Semel} M.,    {Rees} D.~E.,  1992, A\&A, 265, 669

\bibitem[\protect\citeauthoryear{{Donati}, {Wade}, {Babel}, {Henrichs}, {de
  Jong} \& {Harries}}{{Donati} et~al.}{2001}]{Donati2001-betaCeph}
{Donati} J.-F.,  {Wade} G.~A.,  {Babel} J.,  {Henrichs} H.~f.,  {de Jong}
  J.~A.,    {Harries} T.~J.,  2001, MNRAS, 326, 1265

\bibitem[\protect\citeauthoryear{ESA}{ESA}{1997}]{Hipparcos1997}
ESA, 1997, {The Hipparcos and Tycho Catalogues}, ESA SP-1200

\bibitem[\protect\citeauthoryear{{Fabricius} \& {Makarov}}{{Fabricius} \&
  {Makarov}}{2000}]{Fabricius2000-Hipp-rered}
{Fabricius} C.,  {Makarov} V.~V.,  2000, A\&A, 356, 141

\bibitem[\protect\citeauthoryear{{Gray}}{{Gray}}{2005}]{Gray2005-Photospheres}
{Gray} D.~F.,  2005, {The Observation and Analysis of Stellar Photospheres},
  3rd edn.
Cambridge University Press, Cambridge, UK

\bibitem[\protect\citeauthoryear{{Grevesse}, {Asplund} \& {Sauval}}{{Grevesse}
  et~al.}{2005}]{Grevesse2005-solar_abun}
{Grevesse} N.,  {Asplund} M.,    {Sauval} A.~J.,  2005, in {Alecian} G.,
  {Richard} O.,   {Vauclair} S.,  eds, Element Stratification in Stars: 40
  Years of Atomic Diffusion Vol.~17 of EAS Pub. Ser., {The New Solar Chemical
  Composition}.
p.~21

\bibitem[\protect\citeauthoryear{{Hensberge}, {Pavlovski} \&
  {Verschueren}}{{Hensberge} et~al.}{2000}]{Hensberge2000-Age_of_NGC2244}
{Hensberge} H.,  {Pavlovski} K.,    {Verschueren} W.,  2000, A\&A, 358, 553

\bibitem[\protect\citeauthoryear{{Hubrig}, {Sch{\"o}ller} \& {Yudin}}{{Hubrig}
  et~al.}{2004}]{Hubrig2004-HAeBe}
{Hubrig} S.,  {Sch{\"o}ller} M.,    {Yudin} R.~V.,  2004, A\&A, 428, L1

\bibitem[\protect\citeauthoryear{{Jaschek} \& {Jaschek}}{{Jaschek} \&
  {Jaschek}}{1995}]{Jaschek-behavior_abun}
{Jaschek} C.,  {Jaschek} M.,  1995, {The behavior of chemical elements in
  stars}.
Cambridge University Press, Cambridge

\bibitem[\protect\citeauthoryear{{Kochukhov} \& {Piskunov}}{{Kochukhov} \&
  {Piskunov}}{2002}]{Kochukhov2002-MDI-intro2}
{Kochukhov} O.,  {Piskunov} N.,  2002, A\&A, 388, 868

\bibitem[\protect\citeauthoryear{{Kochukhov}, {Piskunov}, {Ilyin}, {Ilyina} \&
  {Tuominen}}{{Kochukhov} et~al.}{2002}]{Kochukhov2002-MDI_alpha2CVn}
{Kochukhov} O.,  {Piskunov} N.,  {Ilyin} I.,  {Ilyina} S.,    {Tuominen} I.,
  2002, A\&A, 389, 420

\bibitem[\protect\citeauthoryear{{Kupka}, {Piskunov}, {Ryabchikova}, {Stempels}
  \& {Weiss}}{{Kupka} et~al.}{1999}]{Kupka1999-VALD}
{Kupka} F.,  {Piskunov} N.,  {Ryabchikova} T.~A.,  {Stempels} H.~C.,    {Weiss}
  W.~W.,  1999, A\&AS, 138, 119

\bibitem[\protect\citeauthoryear{Kurucz}{Kurucz}{1993}]{Kurucz1993-ATLAS9etc}
Kurucz R.,  1993, {CDROM Model Distribution}, Smithsonian Astrophys. Obs.

\bibitem[\protect\citeauthoryear{{Landstreet}, {Bagnulo}, {Andretta},
  {Fossati}, {Mason}, {Silaj} \& {Wade}}{{Landstreet}
  et~al.}{2007}]{Landstreet2007-surveyFORS1}
{Landstreet} J.~D.,  {Bagnulo} S.,  {Andretta} V.,  {Fossati} L.,  {Mason} E.,
  {Silaj} J.,    {Wade} G.~A.,  2007, A\&A, 470, 685

\bibitem[\protect\citeauthoryear{{Lueftinger}, {Kuschnig}, {Piskunov} \&
  {Weiss}}{{Lueftinger} et~al.}{2003}]{Lueftinger2003-DI_epUMa}
{Lueftinger} T.,  {Kuschnig} R.,  {Piskunov} N.~E.,    {Weiss} W.~W.,  2003,
  A\&A, 406, 1033

\bibitem[\protect\citeauthoryear{{Michaud}}{{Michaud}}{1970}]{Michaud1970-diff%
usion}
{Michaud} G.,  1970, ApJ, 160, 641

\bibitem[\protect\citeauthoryear{{Michaud}, {Charland} \&
  {Megessier}}{{Michaud} et~al.}{1981}]{Michaud1981-diffusion_magneticApBp}
{Michaud} G.,  {Charland} Y.,    {Megessier} C.,  1981, A\&A, 103, 244

\bibitem[\protect\citeauthoryear{{Morel}}{{Morel}}{1997}]{Morel1997-CESAM}
{Morel} P.,  1997, A\&AS, 124, 597

\bibitem[\protect\citeauthoryear{{Oudmaijer}, {van der Veen}, {Waters},
  {Trams}, {Waelkens} \& {Engelsman}}{{Oudmaijer}
  et~al.}{1992}]{Oudmaijer1992-IRAS+HD72106}
{Oudmaijer} R.~D.,  {van der Veen} W.~E.~C.~J.,  {Waters} L.~B.~F.~M.,  {Trams}
  N.~R.,  {Waelkens} C.,    {Engelsman} E.,  1992, A\&AS, 96, 625

\bibitem[\protect\citeauthoryear{{Palla} \& {Stahler}}{{Palla} \&
  {Stahler}}{1993}]{Palla1993-PMS-Evol}
{Palla} F.,  {Stahler} S.~W.,  1993, ApJ, 418, 414

\bibitem[\protect\citeauthoryear{{Piskunov} \& {Kochukhov}}{{Piskunov} \&
  {Kochukhov}}{2002}]{Piskunov2002-MDI-intro1}
{Piskunov} N.,  {Kochukhov} O.,  2002, A\&A, 381, 736

\bibitem[\protect\citeauthoryear{{Press}, {Teukolsky}, {Vetterling} \&
  {Flannery}}{{Press} et~al.}{1992}]{numerical-recipes-Fortran}
{Press} W.~H.,  {Teukolsky} S.~A.,  {Vetterling} W.~T.,    {Flannery} B.~P.,
  1992, {Numerical Recipes in FORTRAN}, 2nd edn.
Cambridge University Press, Cambridge

\bibitem[\protect\citeauthoryear{{Sch{\"u}tz}, {Meeus} \&
  {Sterzik}}{{Sch{\"u}tz} et~al.}{2005}]{Schutz2005-IR_spec}
{Sch{\"u}tz} O.,  {Meeus} G.,    {Sterzik} M.~F.,  2005, A\&A, 431, 165

\bibitem[\protect\citeauthoryear{{Shorlin}, {Wade}, {Donati}, {Landstreet},
  {Petit}, {Sigut} \& {Strasser}}{{Shorlin} et~al.}{2002}]{Shorlin2002}
{Shorlin} S.~L.~S.,  {Wade} G.~A.,  {Donati} J.-F.,  {Landstreet} J.~D.,
  {Petit} P.,  {Sigut} T.~A.~A.,    {Strasser} S.,  2002, A\&A, 392, 637

\bibitem[\protect\citeauthoryear{{Stibbs}}{{Stibbs}}{1950}]{Stibbs1950-oblique%
_rot}
{Stibbs} D.~W.~N.,  1950, MNRAS, 110, 395

\bibitem[\protect\citeauthoryear{{Tikhonov}}{{Tikhonov}}{1963}]{Tikhonov1963}
{Tikhonov} A.~N.,  1963, Soviet Math. Dokl., 4, 1624

\bibitem[\protect\citeauthoryear{{Torres}, {Quast}, {de La Reza},
  {Gregorio-Hetem} \& {Lepine}}{{Torres}
  et~al.}{1995}]{Torres1995-IRAS_TTauri+HD72106}
{Torres} C.~A.~O.,  {Quast} G.,  {de La Reza} R.,  {Gregorio-Hetem} J.,
  {Lepine} J.~R.~D.,  1995, AJ, 109, 2146

\bibitem[\protect\citeauthoryear{{van den Ancker}, {de Winter} \& {Tjin A
  Djie}}{{van den Ancker} et~al.}{1998}]{van_den_Ancker1998-HAeBe-photo/atmo}
{van den Ancker} M.~E.,  {de Winter} D.,    {Tjin A Djie} H.~R.~E.,  1998,
  A\&A, 330, 145

\bibitem[\protect\citeauthoryear{{van Leeuwen}}{{van
  Leeuwen}}{2007a}]{van_Leeuwen2007-Hipparcos_book}
{van Leeuwen} F.,  2007a, {Hipparcos, the New Reduction of the Raw Data}.
Springer, Dordrecht

\bibitem[\protect\citeauthoryear{{van Leeuwen}}{{van
  Leeuwen}}{2007b}]{van_Leeuwen2007-Hipparcos_validation}
{van Leeuwen} F.,  2007b, A\&A, 474, 653

\bibitem[\protect\citeauthoryear{{Vieira}, {Corradi}, {Alencar}, {Mendes},
  {Torres}, {Quast}, {Guimar{\~a}es} \& {da Silva}}{{Vieira}
  et~al.}{2003}]{Vieira2003-HAeBe_ID}
{Vieira} S.~L.~A.,  {Corradi} W.~J.~B.,  {Alencar} S.~H.~P.,  {Mendes}
  L.~T.~S.,  {Torres} C.~A.~O.,  {Quast} G.~R.,  {Guimar{\~a}es} M.~M.,    {da
  Silva} L.,  2003, AJ, 126, 2971

\bibitem[\protect\citeauthoryear{{Wade}, {Bagnulo}, {Drouin}, {Landstreet} \&
  {Monin}}{{Wade} et~al.}{2007}]{Wade2007-HAeBe_survey}
{Wade} G.~A.,  {Bagnulo} S.,  {Drouin} D.,  {Landstreet} J.~D.,    {Monin} D.,
  2007, MNRAS, 376, 1145

\bibitem[\protect\citeauthoryear{{Wade}, {Donati}, {Landstreet} \&
  {Shorlin}}{{Wade} et~al.}{2000}]{Wade2000-highPrecision-correctBz}
{Wade} G.~A.,  {Donati} J.-F.,  {Landstreet} J.~D.,    {Shorlin} S.~L.~S.,
  2000, MNRAS, 313, 851

\bibitem[\protect\citeauthoryear{{Wade}, {Drouin}, {Bagnulo}, {Landstreet},
  {Mason}, {Silvester}, {Alecian}, {B{\"o}hm}, {Bouret}, {Catala} \&
  {Donati}}{{Wade} et~al.}{2005}]{Wade2005-HAeBe_Discovery}
{Wade} G.~A.,  {Drouin} D.,  {Bagnulo} S.,  {Landstreet} J.~D.,  {Mason} E.,
  {Silvester} J.,  {Alecian} E.,  {B{\"o}hm} T.,  {Bouret} J.-C.,  {Catala} C.,
     {Donati} J.-F.,  2005, A\&A, 442, L31

\end{thebibliography}
\bibliographystyle{mn2e}

\label{lastpage}

\end{document}